\documentclass[12pt]{article}

\usepackage{amsmath}
\usepackage{graphicx}
\usepackage{natbib}
\usepackage{comment}
\usepackage{commath}
\usepackage{url} % not crucial - just used below for the URL 

\newcommand{\blind}{0}

\usepackage[pdfpagemode={UseOutlines},bookmarks=true,bookmarksopen=true,
   bookmarksopenlevel=0,bookmarksnumbered=true,hypertexnames=false,
   colorlinks,linkcolor={blue},citecolor={blue},urlcolor={red},
   pdfstartview={FitV},unicode,breaklinks=true]{hyperref}

\usepackage{algorithm}

\usepackage{algpseudocode}

\usepackage{bm}
\usepackage{amsfonts}
\usepackage{caption}

\DeclareMathOperator*{\argmin}{argmin}
\DeclareMathOperator*{\argmax}{argmax}
\usepackage{setspace}
\usepackage{optidef}
\usepackage{tabularx}
\usepackage{amsthm}
\usepackage{subcaption}
\usepackage[nameinlink]{cleveref}

\usepackage{mathrsfs}

\usepackage{dsfont}

\newtheorem{remark}{Remark}

\makeatletter
\DeclareFontFamily{U}{tipa}{}
\DeclareFontShape{U}{tipa}{m}{n}{<->tipa10}{}
\newcommand{\arc@char}{{\usefont{U}{tipa}{m}{n}\symbol{62}}}%
\newcommand{\arc}[1]{\mathpalette\arc@arc{#1}}
\newcommand{\arc@arc}[2]{%
  \sbox0{$\m@th#1#2$}%
  \vbox{
    \hbox{\resizebox{\wd0}{\height}{\arc@char}}
    \nointerlineskip
    \box0
  }%
}

\begin{document}

\sloppy %To avoid overfull error

\def\spacingset#1{\renewcommand{\baselinestretch}%
{#1}\small\normalsize} \spacingset{1}

%%%%%%%%%%%%%%%%%%%%%%%%%%%%%%%%%%%%%%%%%%%%%%%%%%%%%%%%%%%%%%%%%%%%%%%%%%%%%%

\if0\blind
{
  \title{\bf The Mean Shape under the Relative Curvature Condition}
  \author{Mohsen Taheri\footnote{Mohsen Taheri, Department of Mathematics and Physics, University of Stavanger \\ \phantom{} \hspace{10pt} Email: MohsenTaheriShalmani@gmail.com} \quad , \quad Stephen M. Pizer\footnote{Stephen M. Pizer, Department of Computer Science, University of North Carolina at Chapel Hill}  \quad and \quad J\"orn Schulz\footnote{J\"orn Schulz, Department of Mathematics and Physics, University of Stavanger}}
  \date{}  % Leave the date command empty to remove the date
  \maketitle
} \fi

\if1\blind
{
  \bigskip
  \bigskip
  \bigskip
  \begin{center}
    {\LARGE\bf Title}
\end{center}
  \medskip
} \fi

\bigskip
\begin{abstract}
Guaranteeing that Fréchet means of object populations do not locally self-intersect or are thereby affected is a serious challenge for object representations because the objects' shape space typically includes elements corresponding to geometrically invalid objects. We show how to produce a shape space guaranteeing no local self-intersections for specific but important cases where objects are represented by swept elliptical disks. This representation can model a variety of anatomic objects, such as the colon and hippocampus. Our approach of computing geodesic paths in this shape space enables detailed comparisons of structural variations between groups, such as patients and controls. The guarantee is met by constraining the shape space using the Relative Curvature Condition (RCC) of swept regions. This study introduces the Elliptical Tube Representation (ETRep) framework to provide a systematic approach to ensure valid mean shapes, effectively addressing the challenges of complex non-convex spaces while adhering to the RCC. The ETRep shape space incorporates an intrinsic distance metric defined based on the skeletal coordinate system of the shape space. The proposed methodology is applied to statistical shape analysis, facilitating the development of both global and partial hypothesis testing methods, which were employed to investigate hippocampal structures in early Parkinson's disease.
\end{abstract}

\vspace{90pt}
\noindent%
\vfill

\newpage
\spacingset{1.5} % DON'T change the spacing!

\section{Introduction}\label{sec:Introduction}
Object \textit{shapes} play a crucial role when analyzing medical data, particularly in the characterization of diseases in the human body. Given a certain shape representation, shape analysis is often based on Riemannian statistics, such as the computation of the Fréchet mean and Riemannian variance, within abstract metric spaces \citep{pennec2019riemannian}. These spaces need to be conditioned to respect the inherent geometric structure of the underlying shape representation. This ensures that paths within the space traverse only geometrically valid shapes, with distances along these paths providing the foundation for calculating statistical measures like means and variances.\par
In this work, we construct a shape space from object representations using \textit{elliptical tubes} (e-tubes), providing an effective approximation for a class of swept regions with tubular forms similar to e-tubes, with a focus on defining a proper shape space and computing means. Here, an e-tube is a \textit{swept region} generated by a smooth sequence of slicing planes along a central curve, called the \textit{spine}. Each cross-section (i.e., the intersection of a slicing plane with the region) is an elliptical disk oriented normal to the spine, with the spine passing through the centers of these cross-sections \citep{fang1994extruded,ma2018deforming}.\par
\begin{figure}[ht]
\centering
\boxed{\includegraphics[width=0.60\textwidth]{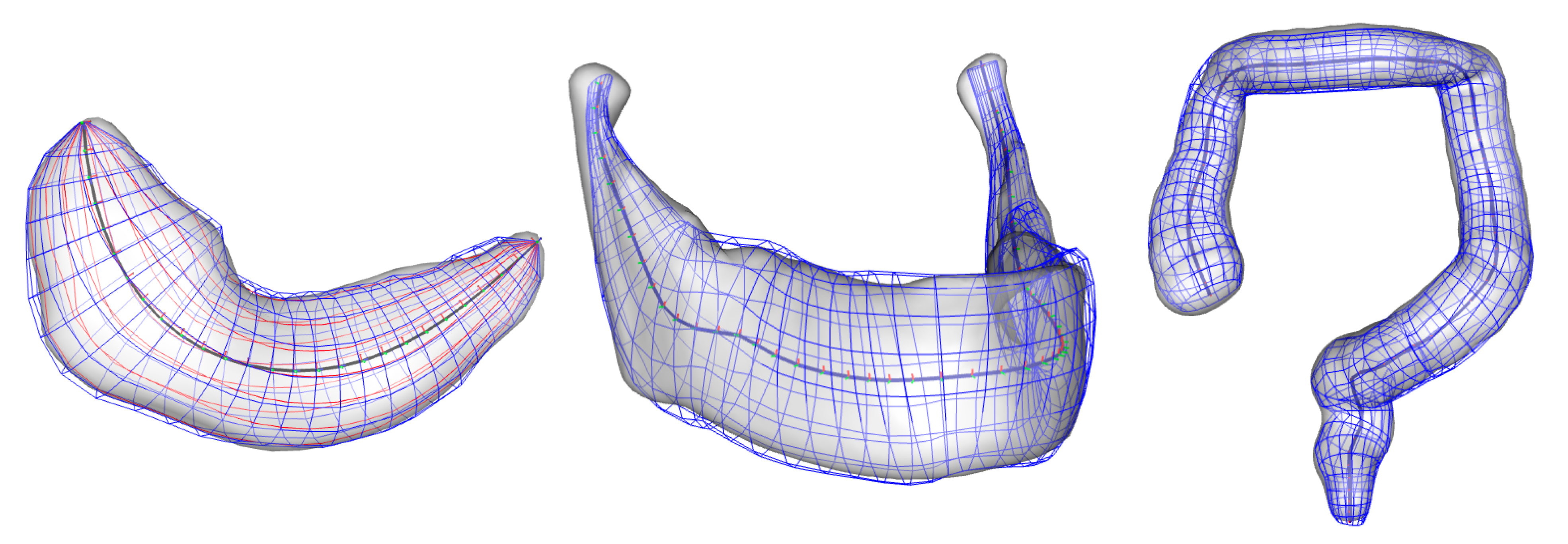}}
\caption{Illustration of a hippocampus, a mandible (excluding the coronoid process), and a colon, shown from left to right. Each object is represented by an e-tube. The e-tubes are modeled as sequences of ellipses along their spines, which are illustrated by dark curves.}
\label{fig:hippoMandibleColon}
\end{figure}
As depicted in \Cref{fig:hippoMandibleColon}, many living organisms, such as vermiform creatures, as well as various human body parts — including the leg bones, intestines, mandible, and most brain subcortical structures like the hippocampus — exhibit forms analogous to e-tubes. These structures often can be adequately approximated and simplified using e-tubes. Calculating the (sample) \textit{mean shape} \citep{dryden2016statistical} for a group of e-tubes is crucial for hypothesis testing and cross-sectional analysis, as it facilitates the identification of underlying patterns and differences between groups.\par
To compute the mean shape, it is essential to define the concepts of shape, shape space, and shape distance. These definitions establish the basis for viewing the mean shape as an element within the shape space. The mean shape is often represented as the Fréchet mean, a point in shape space that minimizes the sum of squared distances to all sample shapes, ensuring a sense in which it is optimally positioned within the sample set \citep{pennec2019riemannian}. However, for the distances to the samples to be valid, the geodesics to the mean must pass only geometrically valid shapes. Therefore, in the scope of e-tube analysis, essential properties of e-tubes must be taken into account. In addition to having elliptical cross-sections normal to the spine, an e-tube, as a swept region with a smooth boundary, must also satisfy the Relative Curvature Condition (RCC) \citep{damon2008swept}. The RCC defines a curvature tolerance that maintains locally non-intersecting cross-sections within the e-tube’s domain, ensuring the smoothness and integrity of the object’s boundary.\par
\begin{figure}[ht]
\centering
\boxed{\includegraphics[width=0.98\textwidth]{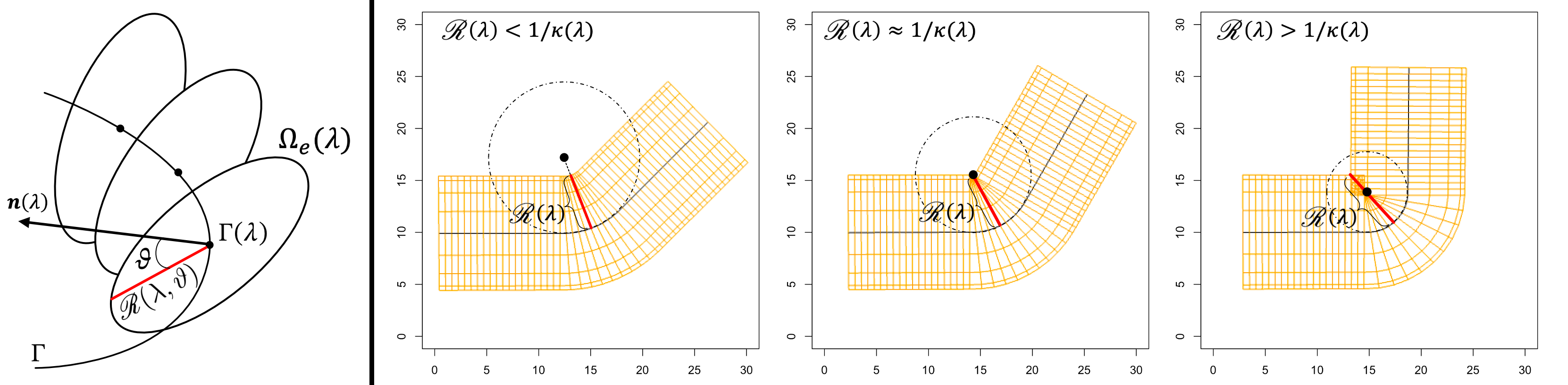}}
\caption{Left: The RCC for an e-tube. Right three figures: The RCC violation in a 2D e-tube by increasing the curvature. The object's width along the normal at the spine's midpoint is $\mathscr{R}(\lambda) \approx 6$ for $\lambda = 0.5$. As the spine's curvature increases the cross-sections begin to intersect within the object as the radius of the osculating circle (i.e., $\frac{1}{\kappa(\lambda)}$) decreases (here approximately from 8 to 6 and then to 4).}
\label{fig:RCC_2D}
\end{figure}

As shown in \Cref{fig:RCC_2D} (left) (and \Cref{sec:e_tubes,sec:discret_e_tubes} discuss in detail), for the e-tube $\Omega_e$ with a parameterized spine $\Gamma(\lambda)$, where ${\lambda\in[0, 1]}$, consider the cross-section $\Omega_e(\lambda)$, which is orthogonal to the spine at $\Gamma(\lambda)$. The distance from the spine to the cross-section's boundary, at a \textit{deviation angle} $\vartheta$ from the spine’s normal $\bm{n}(\lambda)$, must satisfy the RCC: ${\forall\lambda,\vartheta:\quad\mathscr{R} (\lambda, \vartheta) < \frac{1}{\cos(\vartheta) \kappa(\lambda)}}$ when ${\vartheta\in(-\frac{\pi}{2},\frac{\pi}{2})}$ \citep{ma2018deforming}. Here, $\kappa(\lambda)$ is the curvature of the spine at $\Gamma(\lambda)$ (i.e., the inverse of the osculating circle's radius), while the term ${\cos(\vartheta)\kappa(\lambda)}$ represents the \textit{relative curvature}. In 2D, the RCC simplifies to: ${\mathscr{R}(\lambda)<\frac{1}{\kappa(\lambda)}}$ since $\vartheta = 0$. \Cref{fig:RCC_2D} (right) illustrates a violation of the RCC in a 2D tube: as the spine's curvature increases, the cross-sections begin to intersect within the object, highlighting the geometric constraints the RCC imposes.\par
In the shape space conditioned on the RCC, each element represents an e-tube that complies with the RCC. In this framework, geodesics can be understood as transitions from one element to another within the shape space that results in a smooth transformation of the first object into the second. This means that during the transformation, the cross-sections and spine deform smoothly, such that cross-sections remain elliptical disks normal to the spine and satisfy the RCC. In fact, the RCC sets a boundary condition for the shape space. Ignoring this condition can lead to inaccurate statistical inferences, such as calculating the mean shape, that may not be a suitable representation of the sample. The ultimate objective of this work is to define the shape and shape space for e-tubes while incorporating the RCC. \par
\begin{figure}[ht]
\centering
\boxed{\includegraphics[width=0.90\textwidth]{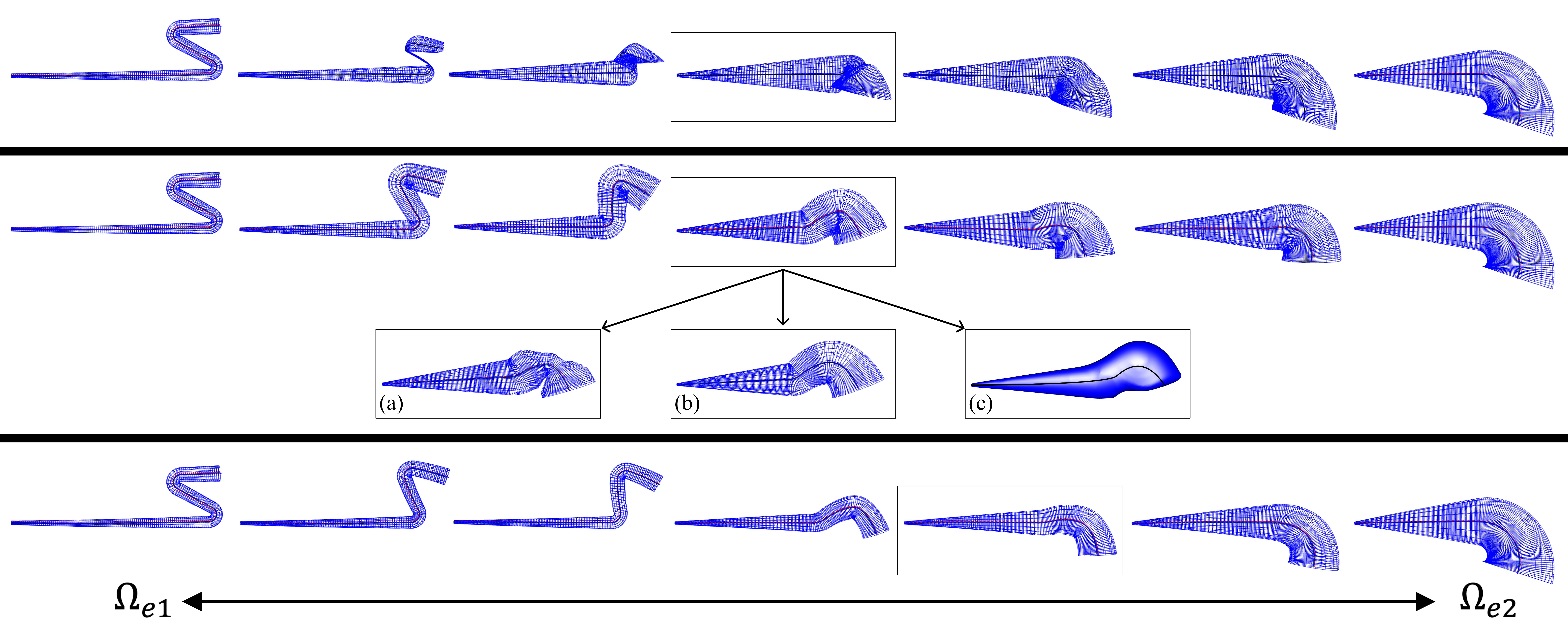}}
\caption{Transformation between two e-tubes, ${\Omega_{e1}}$ and ${\Omega_{e2}}$, with boxed figures indicating the mean shapes. Top row: Collapsed mean shape resulting from Procrustes analysis, illustrating its invalidity. Middle row: Mean shape derived from the non-intrinsic robotic arm transformation, exhibiting a local self-intersection. The mean can be refined to eliminate the self-intersection through (a) shrinkage,\\ (b) spine elongation, and (c) boundary smoothing. Bottom row: Valid transformation and mean shape obtained via the ETRep intrinsic method.}
\label{fig:transformation3D}
\end{figure}

To the best of our knowledge, widely used shape analysis methods such as Procrustes analysis (see \Cref{fig:transformation3D} (top row)) \citep{dryden2016statistical}, functional shape analysis \citep{srivastava2016functional}, elastic shape analysis \citep{jermyn2017elastic}, Euclidean distance matrix analysis \citep{lele2001invariant}, and even skeletal-based techniques like conventional radial distance analysis \citep{thompson2004mapping} and the analysis of medial and skeletal representations \citep{siddiqi2008medial,Taheri2022}—are inadequate for e-tube statistical shape analysis because they do not account for the crucial properties of e-tubes including the RCC.\par
For example, assume a sample of ${m}$ e-tubes, denoted ${\{\Omega_{ej}\}_{j=1}^m}$. In a naive approach for applying Procrustes analysis \citep{dryden2016statistical}, each object can be represented by a distribution of ${n}$ corresponding points across the sample, creating a \textit{point distribution model} (PDM) on the object’s boundary. Procrustes analysis aligns and normalizes the PDMs so that each object can be represented by a unit vector on the hypersphere ${\mathbb{S}^{3n-1}}$ (known as Kendall’s pre-shape space), which is a manifold equipped with the geodesic distance. Consequently, the sample of e-tubes can be viewed as a distribution of $m$ points on the hypersphere. Let the point $\bm{p}_j\in\mathbb{S}^{(3n-1)}$ represent the $j$th object $\Omega_{ej}$. The mean shape corresponding to the Fréchet mean is ${{\bar{\bm{p}}=\argmin_{\bm{p}\in\mathbb{S}^{(3n-1)}}\sum_{j=1}^md_g^2(\bm{p}_j,\bm{p})}}$, where ${d_g(\bm{p}_j,\bm{p})=\cos^{-1}{(\bm{p}_j\cdot\bm{p})}}$ is the geodesic distance.\par
As illustrated in the top row of \Cref{fig:transformation3D}, such PDM approaches may yield Fréchet means that suffer from self-intersections, non-normality, and even collapses because they are derived solely from boundary points, disregarding the essential properties of e-tubes. This issue arises from the transformation path, defined by a geodesic on the hypersphere, which may traverse invalid elements, ultimately rendering the computed mean shape geometrically invalid. Thus, in the context of Procrustes analysis, the e-tube space can be conceptualized as a hypersphere with intricate holes and discontinuities, where hypersphere geodesics that do not account for the structure of the underlying space are non-intrinsic paths that may traverse the holes. This makes analysis within this framework highly challenging.\par
Apparently, to preserve the tubular structure of e-tubes during transformations, we can leverage concepts from robotic arm kinematics, extensively discussed in the literature \citep{murray2017mathematical,ahmadzadeh2018trajectory}. Within this framework (as \Cref{sec:Size_and_shape_nonIntrinsic} discusses), e-tubes are represented in a discrete format, where the discrete spine serves as the skeleton of a robotic arm equipped with a finite sequence of local frames (see \Cref{fig:materialFramesHippo}). Each frame defines the orientation of a cross-section. Adjusting the orientation of these frames induces deformation in the object while preserving its tubular structure and central curve throughout the transformation, as depicted in the middle row of \Cref{fig:transformation3D}. To incorporate the RCC and prevent self-intersection, \citet{ma2018deforming} proposed using robotic arm transformations, followed by a refinement process to eliminate self-intersections from invalid objects. In this context, the mean can be treated as an extrinsic mean, determined based on non-intrinsic paths between samples defined by robotic arm transformations and subsequently projection to a location close to the obtained non-intrinsic mean, but within the underlying space. \par
%to a location near, but still within, the underlying space.\par
Regardless of the fact that refinement processes can introduce abrupt changes in transforming objects, leading to non-smooth transformations, Ma's method and similar approaches are fundamentally unacceptable for calculating the mean and making statistical inferences due to the arbitrary nature of the refinement process (which can be applied through various methods), as shown in the middle row of \Cref{fig:transformation3D}. Moreover, even robust extrinsic methods in non-convex spaces that try to consider the extrinsic mean as the closest element of the space to the pre-projected mean are problematic in non-convex spaces, as the closest element may not be unique. Such extrinsic means, which typically lie on the space's boundary, fail to adequately represent the interior sample set, as illustrated by an example in the second section of the \hyperlink{link_Supplementary}{\textit{Supplementary Materials}(Supp.Mat.)}.\par
The literature lacks a comprehensive discussion on the concepts of shape and shape space for e-tubes. Additionally, it does not adequately address intrinsic distances for computing means, which require careful consideration of the space's structure and its intrinsic properties. To fill this gap, we introduce the \textit{e-tube representation} (ETRep), a robust, alignment-independent framework for representing e-tubes that is invariant to rigid transformations. This alignment independence allows for consistent and unbiased statistical shape analysis \citep{pizer2022skeletons, Taheri2022}.\par
The ETRep space is structured as the product of the spaces of cross-sections, with each space interpreted as a stratum of high-dimensional non-convex trumpet-like swept regions, which we refer to as a \textit{hypertrumpet}. The intrinsic distance is defined using the \textit{skeletal coordinate system} of the shape space (see \Cref{sec:ETRep_Analysis}). This approach provides a method for deforming e-tubes while maintaining the RCC, as shown in the bottom row of \Cref{fig:transformation3D}.\par
To complement this study, we have developed an R package ``ETRep'' \citep{ETRep2024}, built upon the methodologies presented here. This package provides significant tools for calculating mean shapes and performing e-tube transformations based on the defined shape space, supporting comprehensive statistical shape analysis on e-tubes.\par
The structure of this work is as follows. \Cref{sec:BasicTermsAndDefinitions} of this paper discusses basic terms and definitions related to the swept regions with tubular structures. \Cref{sec:e_tubes} introduces e-tubes, describing their properties, including the structure of local frames along the spine. \Cref{sec:discret_e_tubes} explores e-tubes in a discrete form, enabling computational processing, defining correspondences across samples, and establishing smooth e-tube transformations. That section also defines the RCC for discrete e-tubes, which is crucial for constructing the ETRep and ETRep space. Assuming that a discrete e-tube is a valid discretization of a continuous e-tube, \Cref{sec:ETRep} presents the ETRep as an invariant e-tube representation by defining a locally parameterized spine constructed from local frames. \Cref{sec:ETRep_Analysis} details the ETRep analysis using an intrinsic method that incorporates the RCC and a non-intrinsic method (for comparison) based on the robotic arm transformation that disregards the RCC. The intrinsic approach is grounded in the intrinsic skeletal coordinate system of the hypertrumpet and serves as the foundation for the desired method of computing the mean, which is the central objective of this work. \Cref{sec:Application} presents the analysis of ETReps, applying hypothesis testing to a real dataset to compare hippocampal differences between patients with Parkinson's disease and a control group. The data is from the ParkWest study \citep{alves2009incidence}, provided courtesy of Stavanger University Hospital and Haukeland University Hospital. Finally, \Cref{sec:Conclusion} summarizes and concludes the work.

\section{Basic terms and definitions}\label{sec:BasicTermsAndDefinitions}
This section provides an overview of the fundamental concepts related to regions with tubular forms, which are essential for defining the ETRep and its corresponding space. We denote by lowercase bold letters vectors or points, i.e., the elements of the $d$-dimensional ($d$D) Euclidean space $\mathbb{R}^{d>1}$.\par
The set ${\Omega\subset\mathbb{R}^{d}}$ is a ${d}$D object if it is homeomorphic to a ${d}$D closed ball, where a ${d}$D closed ball is a set ${{B_{r}^d(\bm{p})=\{\bm{x}\in\mathbb{R}^d\mid\norm{\bm{x}-\bm{p}}\leq{r}\}}}$ with center ${\bm{p}\in\mathbb{R}^d}$ and radius ${r\in\mathbb{R}^+}$. A point ${\bm{p}\in\Omega}$ is an interior point of ${\Omega}$ if ${\exists{r}\in\mathbb{R}^+}$ such that ${B^d_r(\bm{p})\subset\Omega}$. Let ${\Omega_{in}}$ be the set of all interior points of ${\Omega}$. Then, the boundary of ${\Omega}$ is ${\partial\Omega=\Omega/\Omega_{in}}$. We say that the boundary $\partial\Omega$ is \textit{smooth} if there exists a diffeomorphic mapping between $\partial\Omega$ and $\mathbb{S}^{d-1}$, where $\mathbb{S}^{d-1}$ denotes the $(d-1)$-sphere ${\mathbb{S}^{d-1}=\{\bm{x}\in\mathbb{R}^d\mid\norm{\bm{x}}=1\}}$ representing the surface of the unit ball ${\mathbb{B}^d}={B_{1}^d(\bm{0})}$ \citep{siddiqi2008medial,jermyn2017elastic}.\par
\begin{figure}[ht]
\centering
\boxed{\includegraphics[width=0.45\textwidth]{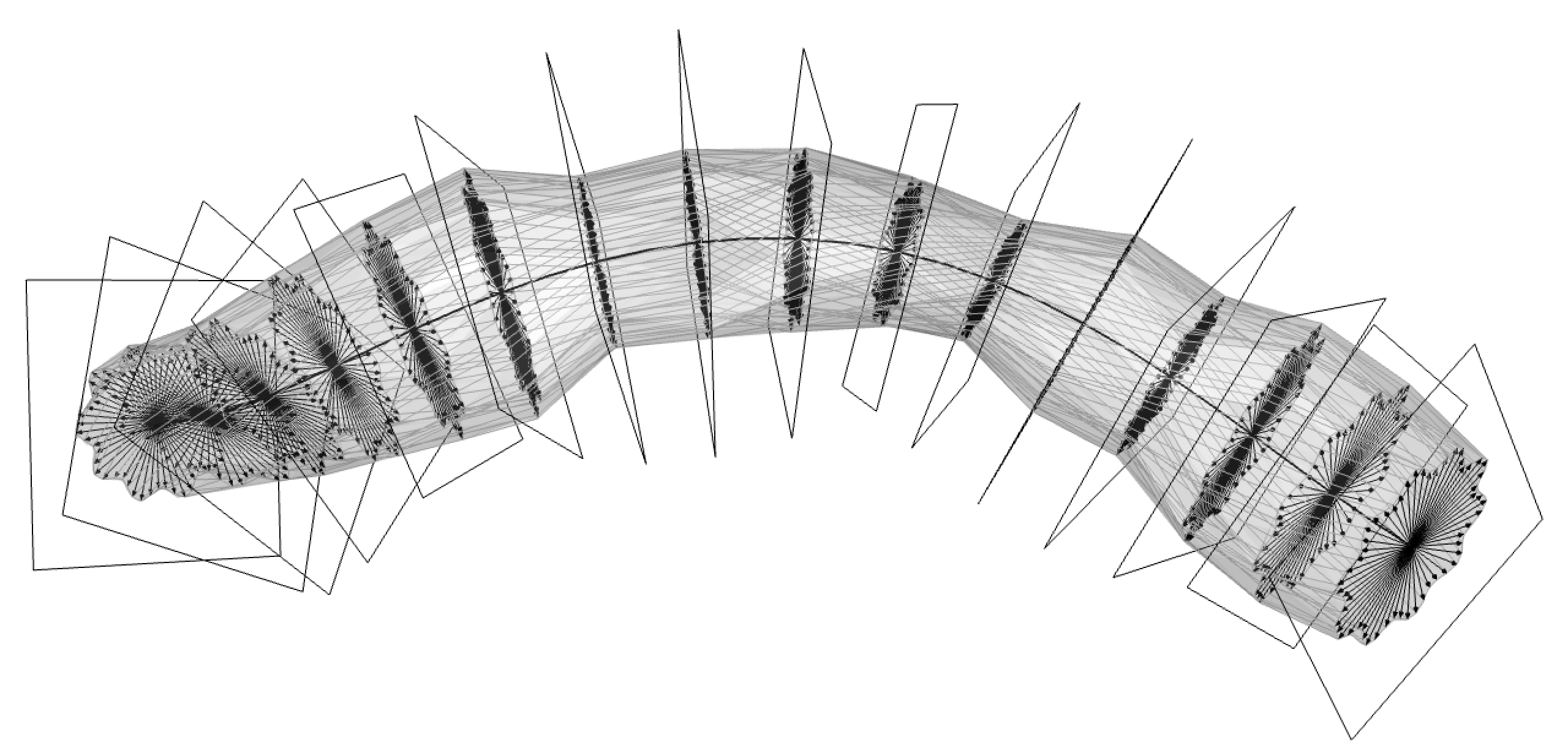}}
\caption{Illustration of a 3D generalized tube with slicing planes and star-convex cross-sections along the spine. Arrows represent the radial vectors.}
\label{fig:starConvexTubular}
\end{figure}
A 1D finite smooth curve embedded in $\mathbb{R}^d$ (like the spine of an e-tube) is the image of a diffeomorphic mapping from a real parameter $\lambda \in [0, 1]$ to $\mathbb{R}^d$, denoted by $\Gamma(\lambda): [0,1] \to \mathbb{R}^d$, which defines the trajectory of the curve and ${\Gamma(0)}$ and ${\Gamma(1)}$ denote the curves' endpoints. Assume ${\Gamma{(\lambda)}}$ as a smooth interior curve of a $d$D object $\Omega$ (i.e., $\forall \lambda$: $\Gamma(\lambda)\in\Omega$), and let ${\Pi(\lambda)}$ denote a ${(d-1)}$D plane\footnote{A ${(d-1)}$D plane crossing the point $\bm{p}$ with the normal ${\bm{n}}$ is the set ${\{\bm{x}\in\mathbb{R}^d\mid\bm{n}\cdot(\bm{x}-\bm{p})=0\}}$. } normal to ${\Gamma}$ at ${\Gamma(\lambda)}$. The $\Omega$ is a swept region with smooth boundary based on $\Gamma$ if it is a disjoint union of cross-sections $\Omega(\lambda)=\Omega\cap\Pi(\lambda)$ (i.e., any point of $\Omega$ lies in exactly one slicing plane) and for each $\lambda$, the cross-section $\Omega(\lambda)$ is a $(d-1)$D object with a smooth boundary \citep{damon2008swept}. In this work, the swept region $\Omega$ is a tube if its cross-sections are closed balls centered along the ${\Gamma}$. A $d$D right tube (or a cylinder) with a straight spine and identical cross-sections as unit balls can be represented as the product space ${\mathbb{B}^{d-1}\times[0,L]}$ (or equivalently ${[0,1]\times\mathbb{S}^{d-2}\times[0,L]}$), where ${L\in\mathbb{R}^+}$ is the spine's length. We define a \textit{unit tube} as a right tube with $L=1$, and we assume a right tube is an \textit{infinite tube} when ${L\rightarrow\infty}$.\par
We consider $\Omega$ as a \textit{generalized tube} or a \textit{generalized cylinder} if $\forall{\lambda}$ the $\Omega(\lambda)$ is star-convex with the center $\Gamma(\lambda)$ (i.e., at $\Gamma(\lambda)$, symmetry is maximized within $\Omega(\lambda)$, and for any point in $\Omega(\lambda)$, the line segment connecting the point to $\Gamma(\lambda)$ lies within $\Omega(\lambda)$). Thus, $\Gamma$ serves as a central curve, referred to as the spine \citep{ballard1982representations}. In this sense, e-tubes are specific cases of 3D generalized tubes whose cross-sections are eccentric elliptical disks \footnote{An eccentric elliptical disk with center $\bm{p}$ and principal radii $a>b$ embedded in $\mathbb{R}^3$ is the set of all points $\bm{x} \in \mathbb{R}^3$ such that $\bm{x}$ lies on the plane $\bm{n} \cdot (\bm{x} - \bm{p}) = 0$ and satisfies the inequality $\frac{(\bm{a} \cdot (\bm{x} - \bm{p}))^2}{a^2} + \frac{(\bm{b} \cdot (\bm{x} - \bm{p}))^2}{b^2} \leq 1$, where $\bm{n}$ is the normal of the plane, and $\bm{a}$ and $\bm{b}$ are orthogonal principal unit axes.} and whose spine passes through the centers of the elliptical cross-sections.\par
Let $\Omega$ be a generalized tube. Any line segment within $\Omega(\lambda)$ that connects $\Gamma(\lambda)$ to a boundary point in $\partial\Omega(\lambda)$ can be seen as a vector with the tail at $\Gamma(\lambda)$, which we refer to as a \textit{radial vector}. Thus, for a generalized tube, there is a field of radial vectors along the spine, where their tips collectively form the object’s boundary while their tails constitute the spine, as depicted in \Cref{fig:starConvexTubular}. This structure is considered as the \textit{swept skeletal structure} of the region, conceptualized as a vector bundle with the spine serving as its base space \citep{damon2008swept}. The designation reflects that radial vectors with a common tail are coplanar and the slicing planes sweep the vector bundle. \Cref{sec:size_and_shape_analysis_Intrinsic} uses the swept skeletal structure's properties to outline the shape space for intrinsic mean calculation. For an e-tube, the radial vectors are along the cross-sections' (polar) radii. \par
In the following section, we present an overview of generalized tubes that can be represented using e-tubes. Additionally, we highlight key properties of e-tubes that are essential for defining both the concept of e-tube shape and the associated shape space.
\section{Elliptical tubes}\label{sec:e_tubes}
\begin{figure}[ht]
\centering
\boxed{\includegraphics[width=0.98\textwidth]{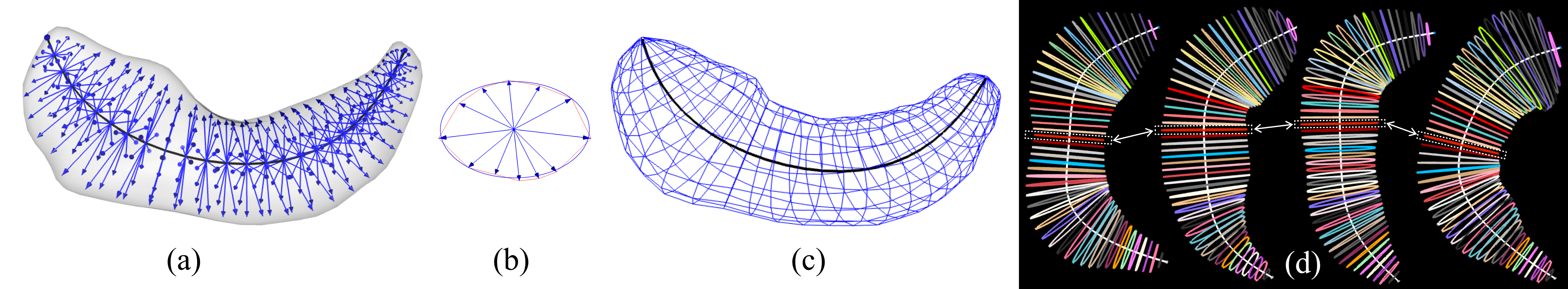}}
\caption{(a) The swept skeletal structure of a left hippocampus as a 3D generalized tube. The arrows are radial vectors. (b) A cross-section is approximated by an elliptical disk. (c) The approximating e-tube of the object is formed by its elliptical cross-sections. (d) Four hippocampi are represented as e-tubes, based on 53 corresponding cross-sections. Double-sided arrows link the 25th corresponding cross-sections.}
\label{fig:hippoAsSweptRegion}
\end{figure}
As defined in \Cref{sec:BasicTermsAndDefinitions}, an e-tube is a 3D generalized tube whose cross-sections are eccentric elliptical disks. A 3D generalized tube ${\Omega}$ can be simplified and represented by an e-tube if there is an e-tube ${\Omega_e}$ with the spine $\Gamma$ such that $\Gamma$ can be considered as the spine of the ${\Omega}$, and ${\forall\lambda \in [0,1]}$, the area of ${\Omega_e(\lambda)}$ closely approximates the area of ${\Omega(\lambda)}$. This approximation can be quantified by the Jaccard index ${{J(\Omega(\lambda),\Omega_e(\lambda))=\frac{|\Omega(\lambda)\cap \Omega_e(\lambda)|_A}{|\Omega(\lambda)\cup\Omega_e(\lambda)|_A}\approx 1}}$, where ${|.|_A}$ denotes the area measurement. Such simplifications are highly valuable for cross-sectional and volumetric analysis of tubular objects that can be seen as generalized tubes, as they allow for easy comparison of the corresponding cross-sections. \Cref{fig:hippoAsSweptRegion} (a-d) present: (a) the swept skeletal structure of a left hippocampus represented as a 3D generalized tube; (b) an ellipse approximating the boundary of a cross-section; (c) the resulting e-tube formed by elliptical cross-sections; and (d) the representation of four distinct hippocampi modeled as e-tubes, derived from 53 corresponding cross-sections. These representations serve as the basis for analyzing the ParkWest data of \Cref{sec:Application}. In \hyperlink{link_Supplementary}{Supp.Mat.}, a brief discussion on e-tube model fitting is provided.\par
A notable property of e-tubes stems directly from their definition. By definition, an e-tube consists of a smooth sequence of elliptical cross-sections, each with a major axis represented as a line segment. The concatenation of these major axes forms a developable surface (i.e., a ruled surface that can be swept out by moving a line), which we refer to as the \textit{major-axes sheet} with two sides, namely the \textit{positive} and the \textit{negative} sides. Since the spine passes through the centers of the cross-sections, it lies on the major-axes sheet. Thus, there exists a moving Darboux frame \citep{linn2020discrete} with one component tangent to the spine and one component normal to the sheet (along the cross-sections' minor axes), which we consider as the (principal) \textit{material frame} of the sheet. The material frame acts like the track frame of a roller coaster, twisting and turning as it travels along the spine of the major-axis sheet. The orientation changes of such a frame can be described by the three Tait-Bryan angles—\textit{roll}, \textit{pitch}, and \textit{yaw}—as outlined in \hyperlink{link_Supplementary}{Supp.Mat}.\par
\begin{figure}[ht]
\centering
\boxed{\includegraphics[width=0.60\textwidth]{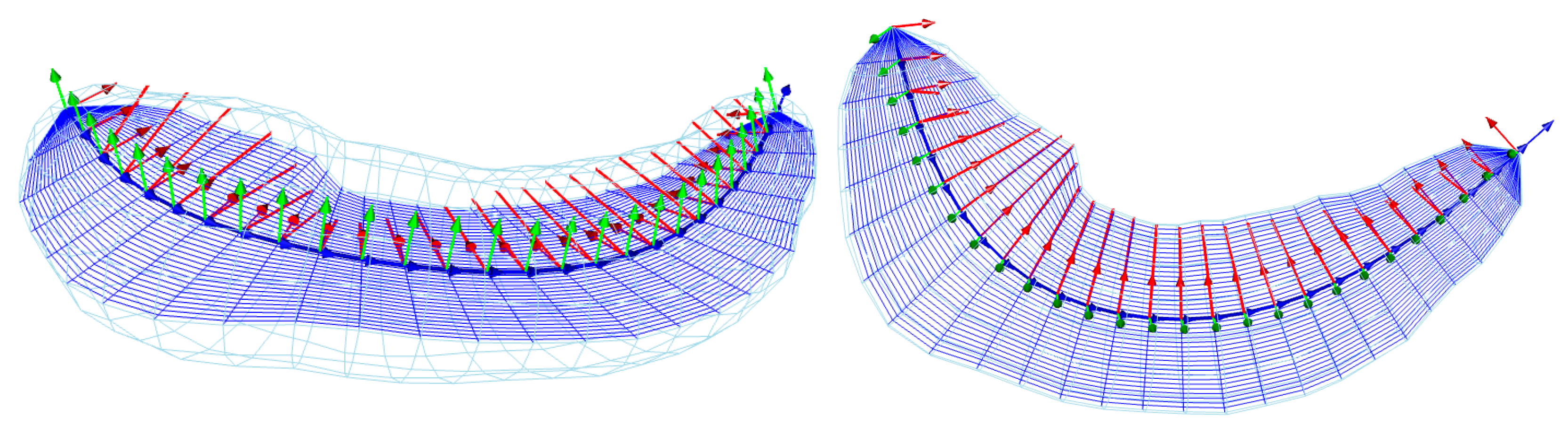}}
\caption{An e-tube representation of a hippocampus. The major-axes sheet is formed by the union of the cross-sections' major axes. Material frames are illustrated by arrows positioned along the spine on the major-axes sheet. Line segments extending from the origins of the frames indicate the spine's normals.}
\label{fig:materialFramesHippo}
\end{figure}
The material frame offers a distinct advantage over frames that focus solely on the geometry of the spine, such as the \textit{Frenet frame}\footnote{For a 1D curve embedded in $\mathbb{R}^d$, the (generalized) Frenet frame consists of a sequence of ${d}$ mutually orthogonal vectors describing the curve's geometric properties. The first vector is the tangent vector. Subsequent vectors, known as normals, are orthogonal to the tangent and to each other.} and its derivatives, like the \textit{rotation-minimizing frame}\footnote{A rotation-minimizing frame is a modified Frenet frame along a curve where the tangent vector remains the same, but the other vectors are adjusted to minimize unnecessary rotation \citep{farouki2016rational}.}. The key strength of the material frame lies in its ability to capture critical information about the twist of the e-tubes by integrating the geometry of both the spine and the major-axes sheet, thereby providing a more comprehensive understanding of the e-tube structure. As a result, we adopt the material frame as the preferred framework for ETRep definition and transformations, as \Cref{sec:ETRep,sec:ETRep_Analysis} discuss.\par
Assuming $\Gamma(\lambda)$ as the spine of an e-tube, for each $\lambda$, the material frame is defined as ${F(\lambda)=(\bm{t}(\lambda),\bm{a}(\lambda),\bm{b}(\lambda))\in SO(3)}$, where $\bm{t}(\lambda)$ is the spine's unit tangent at $\Gamma(\lambda)$, ${\bm{a}(\lambda)}$ and ${\bm{b}(\lambda)}$ are unit vectors aligned with the semi-major and semi-minor axes of the elliptical cross-section at each ${\lambda}$, and $SO(3)$ denotes the rotation group (i.e., ${SO(d)=\{F\in\mathbb{R}^{d\times d}\, | \, F^\top F = \text{diag}(1)_{d \times d} \text{ and } \det(F) = 1 \}}$). We select ${F(\lambda)}$ such that ${\forall \lambda: \bm{b}(\lambda)}$ lies on the positive side of the sheet. Thus, based on the right-hand rule, the orientation induced by the material frame is consistent with the orientation of the sheet. \Cref{fig:materialFramesHippo} illustrates the material frames, the major-axes sheet, the spine, and the spine's normals of an e-tube representing a hippocampus.\par
Let ${a(\lambda)>b(\lambda)}$ be positive scalars representing the principal radii of the cross-section $\Omega_e(\lambda)$. The eccentricity of $\Omega_e(\lambda)$ is given by ${\sqrt{1 -\frac{b(\lambda)^2}{a(\lambda)^2}}}$. Thus, a 2D e-tube can be seen as an e-tube with a flat major-axis sheet, a planar spine, and cross-sectional eccentricity of 1 (i.e., $\forall\lambda$: $a(\lambda)\gg 0$ and $b(\lambda)=0$), as illustrated in \Cref{fig:RCC_2D} (right). In addition, a 3D (circular) tube can be regarded as an e-tube when the cross-sectional eccentricity is infinitesimal (i.e., $\forall\lambda$: $a(\lambda)\approx b(\lambda)$). In this scenario, we assume the material frame coincides with the Frenet frame (or the rotation-minimizing frame).\par
So far, we have discussed e-tubes. However, digital systems inherently operate with discrete representations of data rather than continuous ones. Consequently, converting continuous models into discrete elements facilitates computation and enables smooth transformations, as mentioned in \Cref{sec:Introduction}. Moreover, establishing correspondences between samples of e-tubes is a critical aspect of their analysis. The process of discretization can be achieved for example through curve registration of the spine \citep{srivastava2016functional}. Regardless of the type of registration method, the process aims to establish correspondence between finite sets of corresponding elliptical cross-sections across groups of e-tubes (see \Cref{fig:hippoAsSweptRegion} (d)). The size and orientation of these corresponding cross-sections can then be compared to identify differences such as atrophy, inflammation, elongation, bending, and twisting, as \Cref{sec:Application} explains. Making such comparisons valid crucially requires integrating the RCC into the concept of discrete e-tubes is crucial. Given the variety of methods for defining the curvature of a discrete spine \citep{linn2020discrete}, a careful investigation of the RCC for discrete e-tubes is essential and forms the focus of the following section.

\section{Discrete Elliptical tubes}\label{sec:discret_e_tubes}
In this section, we examine the e-tube properties in discrete format, including the \textit{discrete material frame}. Further, we use these properties to explain the RCC for discrete e-tubes.\par
Let ${\Gamma}$ denote the spine of the e-tube ${\Omega_e}\subset\mathbb{R}^3$. By selecting a finite number ${n\in\mathbb{N}}$ of points on the spine, we define a discrete spine as ${\Gamma(\lambda_i)}$ and ${n}$ corresponding cross-sections as ${\Omega_e(\lambda_i)}$, where ${\lambda_i=\frac{i}{n}}$ for ${i=0,...,n}$. Thus, the discrete e-tube is the subset of ${\Omega_e}$ as ${\bigsqcup_{i=1}^{n}\Omega_e(\lambda_i)}$. The envelope of the cross-sections' boundaries $\partial\Omega_e(\lambda_0)$,...,$\partial\Omega_e(\lambda_n)$ forms a continuous boundary, referred to as the \textit{implied boundary}. It is evident that ${\Omega_e\approx\bigsqcup_{i=1}^{n}\Omega_e(\lambda_i)}$ when ${{n>>1}}$. Therefore, with a sufficiently large number of cross-sections, the implied boundary closely approximates and effectively represents the e-tube’s boundary. For simplicity in writing, we denote cross-section ${\Omega_e(\lambda_i)}$ and its associated slicing plane ${\Pi(\lambda_i)}$ by $\Omega_{e(i)}$ and ${\Pi_{(i)}}$, respectively.\par
Following the discussion of \Cref{sec:e_tubes}, we consider a discrete material frame along the spine as ${F_i = (\bm{t}_i, \bm{a}_i, \bm{b}_i) \in SO(3)}$, where $\bm{a}_i$ and $\bm{b}_i$ are unit vectors aligned with the semi-major and semi-minor axes of $\Omega_{e(i)}$, respectively. The vector $\bm{t}_i = \frac{\bm{p}_{i+1} - \bm{p}_i}{\|\bm{p}_{i+1} - \bm{p}_i\|}$ represents the tangent to the discrete spine, where $\bm{p}_i = \Gamma(\lambda_i)$. Taking into account the spine's normal as ${\bm{n}_i=\frac{\bm{t}_{i-1}\times\bm{t}_{i}}{\norm{\bm{t}_{i-1}\times\bm{t}_{i}}}\times\bm{t}_{i}}$ (coplanar with triangle $\triangle\bm{p}_{i-1}\bm{p}_i\bm{p}_{i+1}$), the discrete Frenet frame can be defined as ${T_i=(\bm{t}_i,\bm{n}_i,\bm{t}_i\times\bm{n}_i)}$ \citep{lu2013discrete}. In case $\bm{t}_{i}=\bm{t}_{i-1}$, we assume $\bm{n}_{i}$ as $\bm{n}_{i-1}$. Therefore, having material frames enables the calculation of the Frenet frames, but the reverse is not true.\par
Up to this point, we have defined a discrete e-tube equipped with a discrete material frame. However, to account for the local self-intersection, the RCC for discrete e-tubes remains to be clarified, as the curvature of a discrete spine lacks a universally accepted definition \citep{linn2020discrete}. Accounting for non-local self-intersections significantly increases the complexity of e-tube analysis. This requires defining transformations that can handle tangled, knot-like objects while avoiding non-local self-intersections. In the \hyperlink{link_Supplementary}{Supp.Mat.}, we provide algorithms for calculating the Fréchet mean for a specific class of e-tubes, referred to as \textit{simply straightenable} e-tubes, based on such transformations. In this work, we focus on local self-intersections that can be studied through the RCC, leaving a detailed examination of non-local self-intersections for future research.\par
\begin{figure}[ht]
\centering
\boxed{\includegraphics[width=0.95\textwidth]{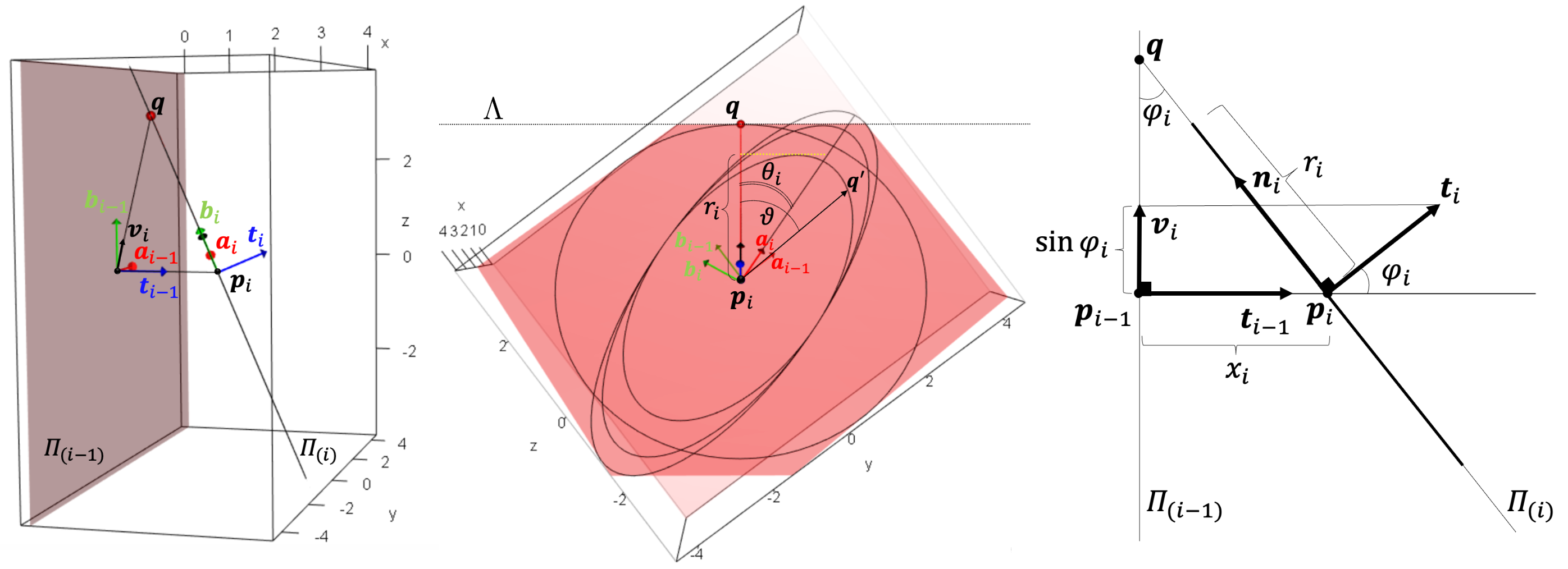}}
\caption{Left: Two consecutive material frames at $\bm{p}_{i-1}$ and $\bm{p}_i$, where ${\bm{t}_{i-1}}$ and ${\bm{t}_i}$ define the normals of non-parallel slicing planes ${\Pi_{(i-1)}}$ and ${\Pi_{(i)}}$. Middle: Elliptical cross-sections of varying sizes with an identical twisting angle $\theta_i$. The circle represents a circular cross-section with zero eccentricity, where $a_i=b_i=\norm{\overrightarrow{\bm{p}_i\bm{q}}}$. Here, the $r_i$ is the magnitude of the projection of the smallest cross-section along the normal, and $\vartheta$ represents the deviation angle of an arbitrary radial vector $\overrightarrow{\bm{p}_i\bm{q}'}$. The greatest cross-section violates the RCC as it crosses the intersecting line $\Lambda$. Right: Illustration of a slice of the left and middle figures crossing ${\bm{p}_{i-1}}$, ${\bm{p}_i}$, and ${\bm{q}}$.}
\label{fig:RCC_twistingFrames}
\end{figure}

Recall that for a continuous e-tube (or any 3D generalized tube), $\forall \lambda$: all radial vectors at $\Gamma(\lambda)$ with deviation angles $\vartheta$ in the range $(-\frac{\pi}{2}, \frac{\pi}{2})$ must satisfy the RCC, which can be expressed as $\mathscr{R}(\lambda, \vartheta)\cos(\vartheta) < \frac{1}{\kappa(\lambda)}$, where $\mathscr{R}(\lambda, \vartheta)$ represents the length of the radial vector, known as the \textit{radial distance}, with deviation angle $\vartheta$ \citep[Ex. 2.12]{damon2008swept}. Let $\kappa(\lambda) \neq 0$. In this case, the osculating circle (with radius $\frac{1}{\kappa(\lambda)}$) locally overlaps with the spine. Since the osculating circle lies in the osculating plane, its neighboring radii, relative to the normal, align with the neighboring slicing planes that are perpendicular to the osculating plane. The intersection of these slicing planes can be seen as a line perpendicular to the osculating circle that passes through its center. The left-hand side of the inequality, $\mathscr{R}(\lambda, \vartheta) \cos(\vartheta)$, represents the length of the projection of the radial vectors onto the spine's normal at $\Gamma(\lambda)$ (see \Cref{fig:RCC_twistingFrames} (middle)). This implies that the radial vectors cannot intersect the neighboring slicing planes, as their projection length onto the normal is less than $\frac{1}{\kappa(\lambda)}$, preventing them from reaching the intersecting line. This is a key restriction imposed by the RCC, demonstrating that, regardless of the size of neighboring cross-sections, a cross-section cannot intersect adjacent slicing planes (which is consistent with the condition that every interior point of a swept region lies on exactly one slicing plane). We leverage this restriction to define the RCC for discrete e-tubes. In this framework, the cross-sections remain constrained within their respective planes and do not cross the neighboring planes.\par
Given the sequential arrangement of slicing planes, it is sufficient to discuss the RCC based on the intersection of each slicing plane with its preceding one. Let $\varphi_i\in[0,\frac{\pi}{2}]$ be the angle between the slicing planes $\Pi_{(i-1)}$ and $\Pi_{(i)}$. Thus, ${\varphi_i=d_g(\bm{t}_i,\bm{t}_{i-1})}$, where $d_g$ is the spherical geodesic distance. Let $\varphi_i \neq 0$, which means $\Pi_{(i-1)} \not\parallel \Pi_{(i)}$. Therefore, $\Pi_{(i-1)}$ and $\Pi_{(i)}$ intersect along the line ${\Lambda=\Pi_{(i-1)}\cap\Pi_{(i)}}$, which is parallel to $\bm{t}_i\times\bm{t}_{i-1}$. Referring to \Cref{fig:RCC_twistingFrames}, let $\bm{q}$ be a point on $\Lambda$ such that $\overrightarrow{\bm{p}_i \bm{q}} \perp \Lambda$. Consequently, $\overrightarrow{\bm{p}_i \bm{q}}$ lies along $\bm{n}_i$, making $\bm{q}$ the closest point on $\Lambda$ to $\bm{p}_i$ (\Cref{fig:RCC_twistingFrames} (middle)). This implies that neither the width of $\Omega_{e(i)}$ in the normal direction nor the magnitude of its projection onto $\bm{n}_i$ can exceed $\|\overrightarrow{\bm{p}_i\bm{q}}\|$, thereby ensuring that $\Omega_{e(i)}$ does not intersect $\Pi_{(i-1)}$.\par
Therefore, we can consider the discrete curvature at $\bm{p}_i$ representing the rate of change in the tangent vectors along the curve and consistent with the RCC as ${\kappa(\lambda_i)= \frac{1}{\norm{\overrightarrow{\bm{p}_i\bm{q}}}} = \frac{\sin \varphi_i}{x_i}}$ (which is a commonly used form of discrete curvature alongside $\frac{\varphi_i}{x_i}$, $\frac{\tan \varphi_i}{x_i}$, $ \frac{2\sin\left(\frac{\varphi_i}{2}\right)}{x_i + x_{i+1}}$, etc. \citep[Remark 4.3]{linn2020discrete}), where $\varphi_i$ represents the (bending or) curvature angle, and ${x_i=\norm{\bm{p}_i-\bm{p}_{i-1}}}$ is the size of the ${\overrightarrow{\bm{p}_{i-1}\bm{p}_i}}$ that refers to the ${i}$th \textit{spinal connection} vector.\par
Let ${r_i}$ be the length of the projection of $\Omega_{e(i)}$ in the direction of normal $\bm{n}_i$, i.e., ${r_i=\mathscr{R}(\lambda_i,\vartheta_{max})\cos(\vartheta_{max})}$, where $\vartheta_{max}={\argmax_{\vartheta\in(-\frac{\pi}{2},\frac{\pi}{2})}\mathscr{R}(\lambda_i,\vartheta)\cos(\vartheta)}$. Hence, the RCC in discrete format can be considered as ${r_i < \frac{x_i}{\sin{\varphi_i}}}$, or equivalently, ${r_i <\frac{x_i}{\norm{\bm{v}_i}}}$, where $\bm{v}_i$ is the orthogonal projection of $\bm{t}_i$ onto $\Pi_{(i-1)}$ (providing the hemispherical coordinates of $\bm{t}_i$ in $F_{i-1}$ based on two elements analogous to polar and azimuthal angles), while the vectors $\bm{t}_i$, $\bm{t}_{i-1}$, $\bm{n}_i$, and $\bm{v}_i$ are coplanar, as depicted in \Cref{fig:RCC_twistingFrames} (left and right). Thus, to verify that $\Omega_{e(i)}$ satisfies the RCC, it is sufficient to calculate $r_i$.\par 
As illustrated in \Cref{fig:RCC_twistingFrames} (middle), let ${{\theta_i\in[-\pi,\pi]}}$ be the \textit{twisting angle} of $\Omega_{e(i)}$ relative to ${\bm{n}_i}$ based on ${R_{\bm{t}_i}(\theta_i)\bm{a}_{i}=\bm{n}_i}$, where ${{R_{\bm{t}}(\theta)=I_3+\sin{\theta}[\bm{t}]_\times+(1-\cos{\theta})(\bm{t}\otimes\bm{t}-I_3)}}$ is the rotation around the Euler axis ${\bm{t}}$ according to the right-hand rule by ${\theta}$ degrees and ${\otimes}$ denotes the outer product. Then, the ${\partial\Omega_{e(i)}}$ is an ellipse that can be parameterized as ${{\partial\Omega_{e(i)}=(a_i\cos{\eta},b_i\sin{\eta})}}$, where ${\eta\in[0,2\pi]}$. By rotating ${\partial\Omega_{e(i)}}$ with ${\theta_i}$ degree of twist clockwise relative to ${F_i}$, we have

\begin{equation*}
\begin{bmatrix}
\cos{\theta_i} & -\sin{\theta_i} \\
\sin{\theta_i} & \cos{\theta_i} 
\end{bmatrix}
\begin{bmatrix}
a_i\cos{\eta}\\
b_i\sin{\eta}
\end{bmatrix}=\begin{bmatrix}
a_i\cos{\eta}\cos{\theta_i}-b_i\sin{\eta}\sin{\theta_i}\\
a_i\cos{\eta}\sin{\theta_i}+b_i\sin{\eta}\cos{\theta_i}
\end{bmatrix}.
\end{equation*}
Thus, ${a_i\cos{\eta}\cos{\theta_i}-b_i\sin{\eta}\sin{\theta_i}}$ is the parameterized projection of ${\partial\Omega_{e(i)}}$ onto $\bm{n}_i$. Assume function ${\mathscr{H}(\eta;a,b,\theta)=a\cos{\eta}\cos{\theta}-b\sin{\eta}\sin{\theta}}$. For given ${a_i}$, ${b_i}$, and ${\theta_i}$, the maximum value of ${\mathscr{H}(\eta;a_i,b_i,\theta_i)}$ defines ${r_i}$. Hence, based on ${\frac{\partial{\mathscr{H}(\eta;a_i,b_i,\theta_i)}}{\partial{\eta}}=0}$ we have
\begin{equation*}
    -a_i\sin{\eta}\cos{\theta_i}-b_i\cos{\eta}\sin{\theta_i}=0\Rightarrow \frac{\sin{\eta}}{\cos{\eta}}=-\frac{b_i}{a_i}\frac{\sin{\theta_i}}{\cos{\theta_i}}\Rightarrow  \eta=\tan^{-1}\left(\frac{-b_i}{a_i}\tan{\theta_i}\right).
\end{equation*}
As a result, the RCC can be formulated as 
\begin{equation}\label{equ:RCC}
    r_i=\left|a_i\cos{\left(\tan^{-1}\left(\frac{-b_i}{a_i}\tan{\theta_i}\right)\right)}\cos{\theta_i}-b_i\sin{\left(\tan^{-1}\left(\frac{-b_i}{a_i}\tan{\theta_i}\right)\right)}\sin{\theta_i}\right|<\frac{x_i}{\norm{\bm{v}_i}}.
\end{equation}
With the RCC in place, we can define the shape and shape space for discrete e-tubes.
\section{Elliptical tube representation}\label{sec:ETRep}
Shape representations that are inherently invariant to rigid transformations and alignments offer significant advantages for statistical shape analysis. This is because the process of alignment can be somewhat arbitrary and may introduce noise into the analysis \citep{lele2001invariant}. Such representations can be constructed using local frames along the spine, as demonstrated by \citet{pizer2022skeletons,Taheri2022}. In this section, we introduce the e-tube representation (ETRep), an alignment-independent shape representation for discrete e-tubes, which leverages the defined discrete material frame.\par
Assume the spine of an e-tube $\Omega_e\subset\mathbb{R}^3$ as a discrete curve $\Gamma(\lambda_i)$ equipped with discrete material frames ${F_0,...,F_n}$ at spinal points ${\bm{p}_0,...,\bm{p}_n}$, as defined in \Cref{sec:discret_e_tubes}. We make the choice of considering ${F_{i-1}}$ as the parent frame of ${F_i}$, where ${i=1,...,n}$, and we assume ${F_0}$ as the parent of itself. Since the spinal connection ${\overrightarrow{\bm{p}_{i-1}\bm{p}_i}}$ that connects frame ${F_i}$ to its parent ${F_{i-1}}$ is along ${\bm{t}_i}$, the spine can be locally parameterized by a sequence of tuples as ${((F^*_i,x_i))_{i=0}^n}$, where ${F_i^*}$ is the ${i}$th frame's orientation based on its parent. Considering that ${\forall i}$: frame ${F_i}$ is invertible and ${F_i^{-1}=F_i^\top}$ (as it belongs to ${SO(3)}$), the orientation of each frame relative to its parent frame can be determined using the transpose of the parent frame. Let ${F^\dagger}$ be the parent of ${\tilde{F}}$, both expressed in the global coordinate system defined by the identity frame ${I_3=\text{diag}(1)_{3 \times 3}}$. We can align ${F^\dagger}$ to ${I_3}$ by ${{(F^\dagger)^\top F^\dagger=I_3}}$. Consequently, ${{\tilde{F}^*=(F^\dagger)^\top \tilde{F}}}$ represents ${\tilde{F}}$ in its parent coordinate system. Further, if ${\tilde{F}^*}$ is the orientation of ${\tilde{F}}$ based on its parent, then ${F^\dagger\tilde{F}^*}$ is the orientation of ${\tilde{F}}$ based on the global coordinate system ${I_3}$.\par
Based on the locally parameterized spine, we can define the ETRep as the sequence ${s=(\omega_i)_{i=0}^n}$, where ${{\omega_i=(F^*_i,x_i,a_i,b_i)}}$ is the representation of the ${i}$th cross-section ${\Omega_{e(i)}}$ with principal radii ${a_i}$ and ${b_i}$. Evidently, the ETRep is invariant to the act of rigid transformations of translation and rotation.\par
A frame can be represented as a 4D unit quaternion vector in ${\mathbb{S}^3}$, which encodes the frame axis and angle of rotation \citep{huynh2009metrics}. Let ${\bm{f}^*_i}$ be the unit quaternion representation of the frame ${F^*_i}$. Without considering the RCC, by definition ${s}$ can be represented as ${((\bm{f}_i^*,x_i,a_i,b_i))_{i=0}^n}$ residing in the product space ${{\mathcal{S}^{n+1}=(\mathbb{S}^3\times(\mathbb{R}^+)^3)^{n+1}}}$, which forms a differentiable manifold (as a product of differentiable manifolds). However, calculating paths on the manifold without considering the RCC can result in an inappropriate mean shape, as discussed in \Cref{sec:Introduction}. Thus, for the ETRep statistical analysis, it is essential to discuss the paths and metrics, while accounting for the RCC.

\section{ETRep statistical analysis}\label{sec:ETRep_Analysis}
To perform statistical analysis, it is essential to define the distance between any two ETReps, representing the length of a path connecting them. Such a distance can be defined cross-sectionally as the sum of squared distances between the corresponding cross-sections. However, since an ETRep is valid only if all its elliptical cross-sections satisfy the RCC, incorporating the RCC restricts the space to $\mathcal{S}^{n+1}_\text{RCC} \coloneqq \mathcal{S}^{n+1}|_\text{RCC}$, which forms a submanifold of $\mathcal{S}^{n+1}$, i.e., $\mathcal{S}^{n+1}_\text{RCC} \subset \mathcal{S}^{n+1}$. Consequently, (similar to Kendall's pre-shape space) it is possible to have two valid ETReps for which the distance function—defining the length of a naive straight path between them—becomes invalid if the path partially lies outside $\mathcal{S}^{n+1}_\text{RCC}$ (as shown in \Cref{fig:transformation3D}).\par
This section explores the calculation of the mean using intrinsic and non-intrinsic paths, where the intrinsic path accounts for the RCC, while the non-intrinsic path disregards it. Our statistical framework builds on the work of \citet{dryden2016statistical} regarding \textit{size-and-shape} and shape analysis, where size-and-shape refers to geometric properties invariant under rigid transformations, while shape refers to properties invariant under both rigid transformations and uniform scaling. For context, we begin with the non-intrinsic approach, which comes from the robotic arm transformation.
\subsection{Size-and-shape analysis}\label{sec:Size-and-shape_analysis}
\subsubsection{Non-intrinsic approach}\label{sec:Size_and_shape_nonIntrinsic}
Assume two valid ETReps ${s_1, s_2\in\mathcal{S}^{n+1}_\text{RCC}}$ as ${s_j=((\bm{f}_{ji}^*,x_{ji},a_{ji},b_{ji}))_{i=0}^{n}}$, where ${j=1,2}$, such that all of their cross-sections satisfy the RCC of \Cref{equ:RCC}. A straight path connecting ${s_1}$ and ${s_2}$ on the manifold ${\mathcal{S}^{n+1}}$ representing the transformation of the robotic arm discussed in \Cref{sec:Introduction}, could be defined as
\begin{equation}\label{equ:straightPath}
\zeta_s(\gamma;s_1,s_2)=\left(\zeta_g(\gamma;\bm{f}^*_{1i},\bm{f}^*_{2i}),\zeta(\gamma;x_{1i},x_{2i}),\zeta(\gamma;a_{1i},a_{2i}),\zeta(\gamma;b_{1i},b_{2i})\right)_{i=0}^n,
\end{equation}
where $\forall\bm{x},\bm{y}\in\mathbb{S}^d$: ${\zeta_g(\gamma;\bm{x},\bm{y})=\frac{1}{\sin(\xi)}[\sin(\xi(1-\gamma))\bm{x}+\sin(\gamma\xi)\bm{y}]}$ is the geodesic path on the unit sphere, where $\xi=\cos^{-1}(\bm{x},\bm{y})=d_g(\bm{x},\bm{y})$, ${\forall\bm{x},\bm{y}\in\mathbb{R}^d}$: ${\zeta(\gamma;\bm{x},\bm{y})=(1-\gamma)\bm{x}+\gamma\bm{y}}$ is a straight path in the Euclidean space, and ${\gamma\in[0,1]}$. The distance between ${s_1}$ and ${s_2}$ (based on the Pythagorean-like law without applying any arbitrary weighting) can be defined as the length of the straight path connecting them as
\begin{equation}\label{equ:non-intrinsic_distance}
     d_{s}(s_1,s_2)=\left(\sum_{i=0}^nd_g^2(\bm{f}^*_{1i},\bm{f}^*_{2i})+\norm{x_{1i}-x_{2i}}^2+\norm{a_{1i}-a_{2i}}^2+\norm{b_{1i}-b_{2i}}^2\right)^\frac{1}{2}.
\end{equation}
Hence, given a set of ${m}$ observations ${s_1,...,s_m\in\mathcal{S}^{n+1}_\text{RCC}}$. The non-intrinsic mean shape is
\begin{equation}\label{equ:non-intrinsic_mean}
\bar{s}_{\text{non.in}}=\argmin_{s\in\mathcal{S}^{n+1}}\sum_{j=1}^{m}d_s^2(s,s_{{j}}).
\end{equation}
We can consider ${\bar{s}_{\text{non.in}} = (\bar{\omega}_i)_{i=0}^n}$, where ${\bar{\omega}_i = (\bar{\bm{f}}^*_i, \bar{x}_i, \bar{a}_i, \bar{b}_i)}$. Here, ${\bar{\bm{f}}^*_i}$ is the Fréchet mean frame of ${\{\bm{f}^*_{ji}\}_{j=1}^m}$ \citep{moakher2002means}. The values ${\bar{x}_i}$, ${\bar{a}_i}$, and ${\bar{b}_i}$ are the arithmetic means of ${\{x_{ji}\}_{j=1}^m}$, ${\{a_{ji}\}_{j=1}^m}$ and ${\{b_{ji}\}_{j=1}^m}$, respectively. Therefore, the mean ETRep represents an e-tube based on the mean of the corresponding cross-sections.\par
\begin{figure}[ht]
\centering
\boxed{\includegraphics[width=0.75\textwidth]{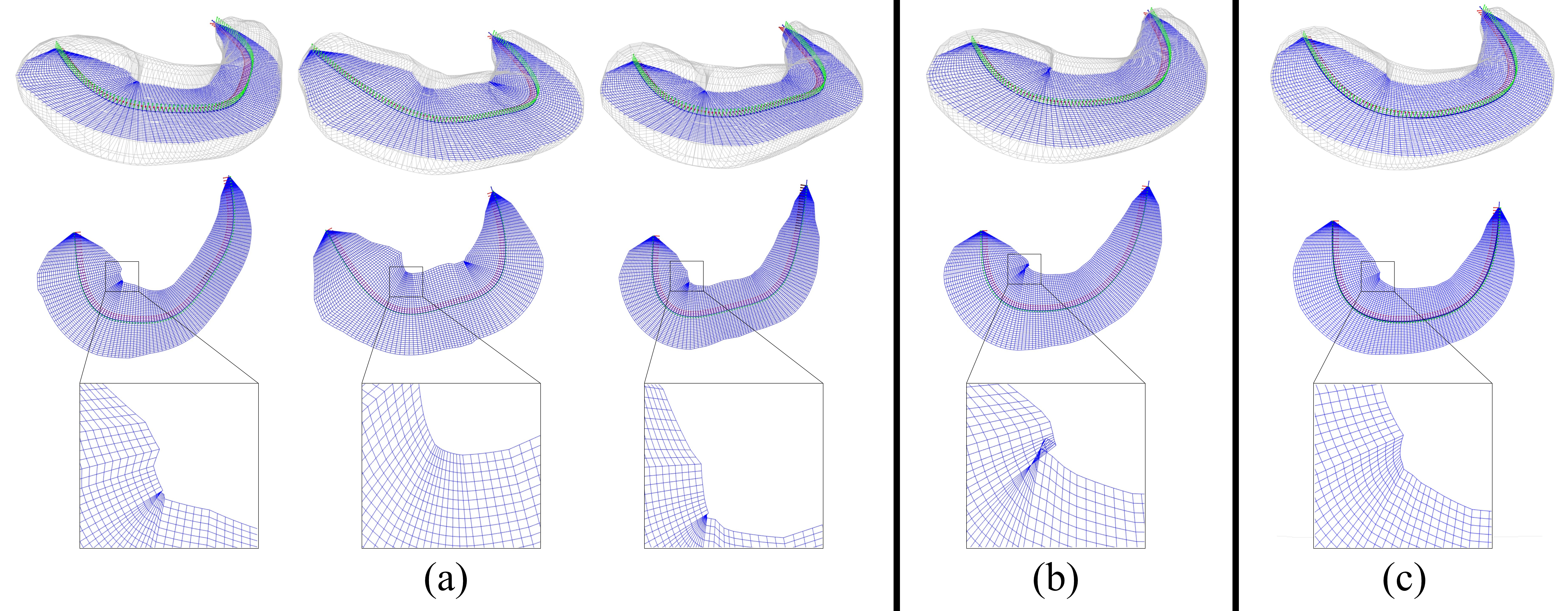}}
\caption{(a) A sample of valid e-tubes representing three hippocampi, with major-axes sheets flattened for clarity. (b) Non-intrinsic mean, showing self-intersection. (c) Intrinsic mean, without self-intersection.}
\label{fig:selfIntersectionHippo}
\end{figure}
Obviously, ${\bar{s}_{\text{non.in}}}$ is an element of ${\mathcal{S}^{n+1}}$, but it is not necessarily an element of ${\mathcal{S}^{n+1}_\text{RCC}}$ (similar to \Cref{fig:IntrinsicMean} (left)). For example, \Cref{fig:selfIntersectionHippo} (b) illustrates the non-intrinsic mean of three ETReps of three hippocampi, showing an evident self-intersection. Such self-intersections highlight the greater complexity of the structure of ${\mathcal{S}^{n+1}_\text{RCC}}$ compared to ${\mathcal{S}^{n+1}}$. This implies that, based on the definition of the non-intrinsic path, the ${\mathcal{S}^{n+1}_\text{RCC}}$ is not a convex space, i.e., ${\exists s_1, s_2\in\mathcal{S}^{n+1}_\text{RCC}}$ and ${\exists\gamma\in(0,1)}$ s.t. ${\zeta_s(\gamma;s_1,s_2)\notin\mathcal{S}^{n+1}_\text{RCC}}$.\par
An initial idea to overcome invalid paths is to define a mapping that transforms the \textit{original space} ${\mathcal{S}^{n+1}_\text{RCC}}$ into a Euclidean convex space, which we refer to as the \textit{transformed space}, ensuring that each straight path in the transformed space corresponds to a valid path in ${\mathcal{S}^{n+1}_\text{RCC}}$. Then, we compute the mean and paths within the transformed space and subsequently map the results back to ${\mathcal{S}^{n+1}_\text{RCC}}$, similar to the approach of \citet{rustamov2009interior}. Although such a mapping offers valid transformations, the choice can be somewhat arbitrary, rendering the resulting mean highly sensitive to that choice, as discussed in Section 4 of \hyperlink{link_Supplementary}{Supp.Mat}.\par
The next section presents an intrinsic (geometry-aware) approach by introducing an alternative ETRep space. This space is equipped with an intrinsic skeletal coordinate system. \Cref{fig:selfIntersectionHippo} (c) and \Cref{fig:transformation3D} (bottom row) illustrate means and transformations based on such an intrinsic approach.\par 
\subsubsection{Intrinsic approach}\label{sec:size_and_shape_analysis_Intrinsic}
We have seen ETReps reside in ${(\mathbb{S}^3\times(\mathbb{R}^+)^3)^{n+1}}$ which is not a homogeneous space, as it is the product of fundamentally different curved and flat components: unit spheres and Euclidean spaces. As a result, defining an appropriate metric for statistical analysis in this space, while incorporating the RCC, poses significant challenges. Thus, we introduce an alternative representation of ETReps within a homogeneous Euclidean space incorporating the RCC. This approach offers a systematic transformation enabling the consistent calculation of the mean shape.\par
% To achieve a homogeneous space, we must combine the information from the material frames with the cross-sections' sizes $a_i$ and $b_i$, ensuring that no information is altered or lost in the process.
Assume an ETRep ${{s=(\omega_i)_{i=0}^n=((F^*_i,x_i,a_i,b_i))_{i=0}^n}}$. The material frame ${F^*_i}$ can be expressed by rotation ${R_{\bm{t}^*_i}(\psi_i)}$, where ${\bm{t}^*_i}$ is the first element of ${F^*_i}$ (defined by the yaw and pitch angles), and ${\psi_i}$ is the roll angle. The (unwrapped) roll angle, representing the cumulative rolling motion, can take distinct values within $ (-\infty, \infty) $, reflecting the extent of clockwise or counterclockwise rolling of ${F^*_i}$ relative to its parent frame. However, we restrict it to $ [-\pi, \pi] $ in this work for practical implementation. As a result, ${\omega_i}$ can be characterized as ${((\bm{t}^*_i,\psi_i),x_i,a_i,b_i)}$ as an element of $(\mathbb{S}^2\times[-\pi,\pi])\times(\mathbb{R}^+)^3$.\par
Furthermore, ${\bm{t}^*_i=(t_{i1},t_{i2},t_{i3})}$ lives on the hemisphere where $t_{i1}>0$ (recall that ${\varphi_i=d_g(\bm{t}_{i-1},\bm{t}_{i})\in[0,\frac{\pi}{2}]}$), so we have ${\bm{v}_i=(t_{i2},t_{i3})}$. Conversely, given ${\bm{v}_i=(v_{i1},v_{i2})}$, we can reconstruct ${\bm{t}^*_i}$ as ${(\sqrt{1-v_{i1}^2-v_{i2}^2},v_{i1},v_{i2})}$. Therefore, given ${\bm{v}_i}$ and ${\psi_i}$, we can compute ${F^*_i}$, and consequently, ${\omega_i}$ can be expressed as a vector ${\bm{\omega}_i=(\bm{v}_i,\psi_i,x_i,a_i,b_i)\in\arc{\mathcal{A}}}$, where ${\arc{\mathcal{A}}=\mathbb{B}^2\times[-\pi,\pi]\times(\mathbb{R}^+)^3\subset\mathbb{R}^6}$ (where $\mathbb{B}^2$ is a 2D unit disk). Similar to the approach of \Cref{sec:Size_and_shape_nonIntrinsic}, we could define the ETRep space as $\arc{\mathcal{A}}^{n+1}$ by disregarding the RCC. In this formulation, paths are considered as straight Euclidean paths, and the resulting mean would be non-intrinsic, aligning closely with the non-intrinsic mean defined in \Cref{equ:non-intrinsic_mean}. We aim to identify a subspace of $\arc{\mathcal{A}}^{n+1}$ that integrates the RCC.\par
To incorporate the RCC, we need to calculate the twisting angle $\theta_i \in [-\pi, \pi]$, which is required for determining $r_i$ (i.e., the magnitude of the projection of $i$th cross-section onto $\bm{n}_i$), as defined in \Cref{equ:RCC}. Given $(\bm{v}_i, \psi_i),$ we can compute the $\theta_i$ because $(\bm{v}_i, \psi_i)$ defines $F^*_i,$ and because $F^*_i$ provides both the normal $\bm{n}_i$ and semi-major axis $\bm{a}_i$ (as $\theta_i$ is the deviation angle of $\bm{a}_i$ from the normal, as shown in \Cref{fig:RCC_twistingFrames} (middle)). Conversely, having $(\bm{v}_i, \theta_i),$ we can determine $(\bm{v}_i, \psi_i),$ as $\bm{v}_i$ provides $\bm{n}_i$ and $\theta_i$ determines $\bm{a}_i$. Thus, $\bm{\omega}_i$ can equivalently be represented by $(\bm{v}_i, \theta_i, x_i, a_i, b_i).$ Since $\theta_i$, $a_i$ and $b_i$ determine the value of $r_i$ in \Cref{equ:RCC}, and $\|\bm{v}_i\|$ ranges within $[0,1],$ the RCC in a compact form can be expressed as $\|\bm{v}_i\| \leq \min\{1, \frac{x_i}{r_i}\}$, where $x_i$ is the $(i-1)$th to $i$th inter-spine point distance. We use this constraint to form the shape space for a cross-section where a valid geodesic path can be formed.\par
\begin{figure}[ht]
\centering
\boxed{\includegraphics[width=0.70\textwidth]{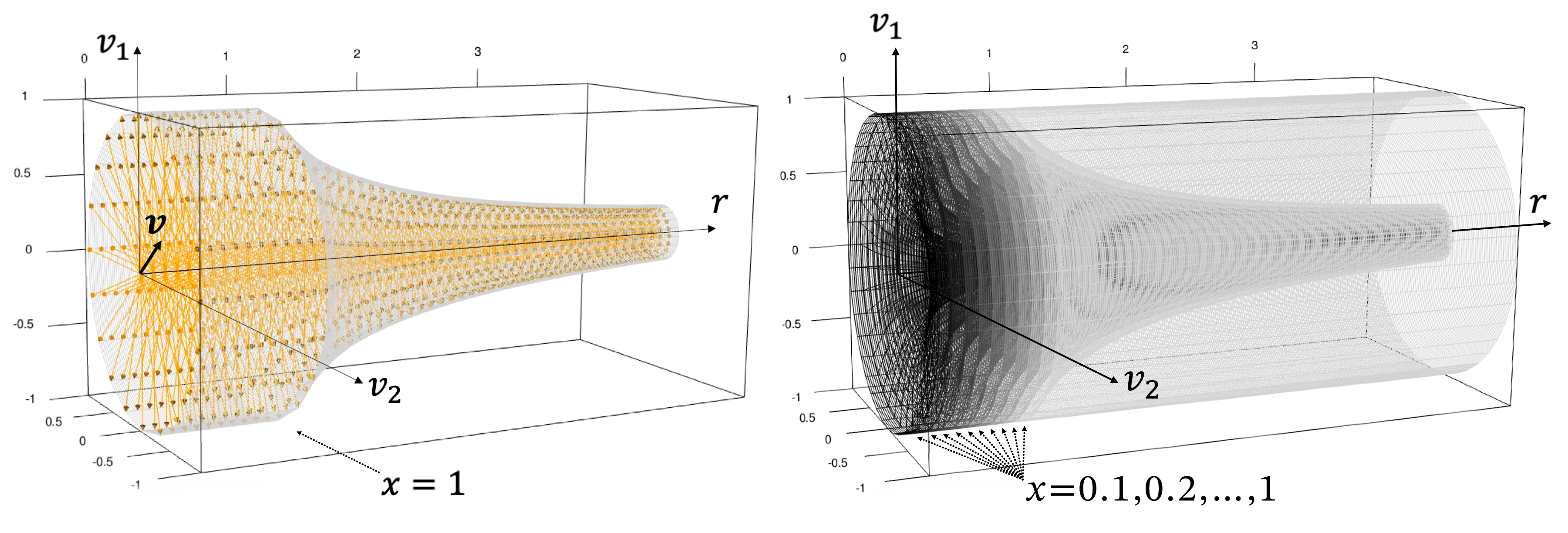}}
\caption{Left: A trumpet as a tube with its swept skeletal structure, representing a slice of $\mathcal{A}_c$ at $x = 1$. Right: Visualization of multiple slices of $\mathcal{A}_c$ for $x = 0.1, 0.2, \dots, 1$, inscribed within the infinite tube.}
\label{fig:hypertrumpet}
\end{figure}

Let ${\bm{\omega}=(\bm{v},\theta,x,a,b)}$ (or $(\bm{v},\psi,x,a,b)$) be an element of $\mathcal{A}$, where ${\mathcal{A}\subset\arc{\mathcal{A}}}$ is the space of valid cross-sections such that $a>b$ and the condition ${\norm{\bm{v}}<\min\{1,\frac{x}{r}\}}$ associated with the RCC is satisfied. From the point of view of the RCC, the 6D space ${\mathcal{A}}$ can be understood as a simplified or compressed 4D space ${\mathcal{A}}_c$ based on the three variables $\bm{v}$, $x$, and $r$. In particular, for a cross-section such as $\bm{\omega}$, the condition $\norm{\bm{v}} < \min\{1, \frac{x}{r}\}$ implies that as $r$ increases, the maximum allowable value of $\norm{\bm{v}}$ decreases in a hyperbolic fashion. Consequently, a slice of $\mathcal{A}_c$ for a locally given $x$ resembles a trumpet-like region bounded by a hyperbolic (pseudospherical) surface, referred to as the \textit{trumpet}. This trumpet can be seen as a straight tube with the $r$-axis as its spine, possessing a swept skeletal structure, as illustrated in \Cref{fig:hypertrumpet} (left). In this context, $\mathcal{A}_c$ can be realized as a layered arrangement of such trumpets over all possible values of $x$. \Cref{fig:hypertrumpet} (right) visualizes $\mathcal{A}_c$ for $x = 0.1, 0.2, \dots, 1$. Based on this realization, we consider ${\mathcal{A}}$ as a hypertrumpet, representing a high-dimensional space that can be conceptualized in a compressed format by a trumpet-like structure.\par
Since 3D (circular) tubes can be viewed as a special case of 3D e-tubes with zero cross-sectional eccentricity (where material frames are assumed to be Frenet frames), the space $\mathcal{A}^{n+1}_c$, by its own, can be interpreted as the space of 3D tubes, with $\mathcal{A}_c$ representing the space of circular cross-sections. This interpretation arises because, under the assumption $\frac{b}{a}=1$, the left-hand side of \Cref{equ:RCC} simplifies to $r=a$ (as ${a(\cos^2\theta + \sin^2\theta) = a}$), where $r$ denotes the radius of the circular cross-sections. Therefore, to enhance the clarity of our approach, we will first discuss the transformed space for 3D tubes, which requires an examination of the skeletal coordinate system for generalized tubes. Then, we will extend this discussion to define the transformed space for 3D e-tubes.\par
\begin{figure}[ht]
\centering
\boxed{\includegraphics[width=0.98\textwidth]{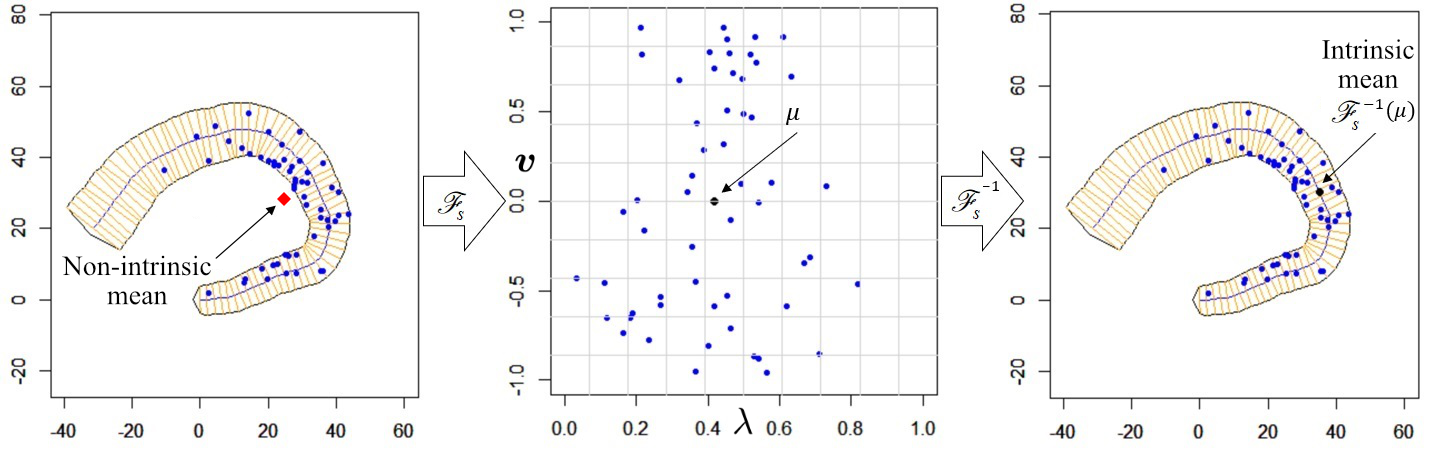}}
\caption{Left: Point distribution within a 2D generalized tube with a swept skeletal structure. The diamond represents the non-intrinsic mean located outside the region. Middle: The mapped distribution in the product space ${[0,1]\times\mathbb{B}^{1}}$, with the Euclidean mean $\mu$ shown inside the space. Right: The intrinsic mean, ${\mathscr{F}_s^{-1}(\mu)}$, depicted as a larger dot, located within the region.}
\label{fig:IntrinsicMean}
\end{figure}

A trumpet's intrinsic coordinate system follows from any generalized tube having an intrinsic coordinate system, which we refer to as the skeletal coordinate system. To clarify the concept, let ${\Omega}$ be a ${d}$D generalized tube with parameterized spine ${\Gamma(\lambda)}$. Thus, ${\lambda\in[0,1]}$ can be seen as the proportion of the curve length at ${\Gamma(\lambda)}$ from the starting point ${\Gamma(0)}$. Assume ${T(\lambda)=(\bm{t}(\lambda),\bm{n}_1(\lambda),\dots,\bm{n}_{d-1}(\lambda))}$ be the (generalized) Frenet frame at ${\Gamma(\lambda)}$, where ${\bm{t}(\lambda)}$ is tangent to the spine and ${\bm{n}_1(\lambda),\dots,\bm{n}_{d-1}(\lambda)}$ span the slicing plane ${\Pi(\lambda)}$. By definition, for any point ${\bm{p}\in \Omega}$, $\exists\lambda\in[0,1] $ such that ${\bm{p}=\Gamma(\lambda)}$ or $\bm{p}$ lies on a unique radial vector if ${\bm{p}\in\Omega\setminus\Gamma}$. Assume that ${\bm{p}\in\Omega\setminus\Gamma}$ lies on a radial vector with tail position $\Gamma(\lambda)\in\mathbb{R}^d$, direction $\bm{u}\in\mathbb{S}^{d-1}$, and length $\mathscr{R}^*(\lambda,\bm{u})\in\mathbb{R}^+$. Thus, ${{\bm{p}=\Gamma(\lambda)+\varsigma\mathscr{R}^*(\lambda,\bm{u})\bm{u}}}$, where ${{\varsigma=\frac{\norm{\overrightarrow{\Gamma(\lambda)\bm{p}}}}{\mathscr{R}^*(\lambda,\bm{u})}\in(0,1]}}$. Let ${(\alpha_1,...,\alpha_d)\in\mathbb{S}^{d-1}}$ represent ${\bm{u}}$ based on the (coordinate system of) ${T(\lambda)}$, i.e., ${{\bm{u}=\alpha_1\bm{t}(\lambda)+\alpha_2\bm{n}_1(\lambda)+...+\alpha_d\bm{n}_{d-1}(\lambda)}}$. Since the radial vector lies on ${\Pi(\lambda)}$, ${\alpha_1=0}$ and we have ${\bm{p}=\Gamma(\lambda)+\varsigma \mathscr{R}^*(\lambda,\bm{u})(0,\alpha_2\bm{n}_1(\lambda),...,\alpha_d\bm{n}_{d-1}(\lambda))}$. Therefore, we can represent ${\bm{p}}$ as ${{(\lambda,\varsigma,(\alpha_2,...,\alpha_d))\in[0,1]^2\times\mathbb{S}^{d-2}}}$. Obviously, each element of the product space ${[0,1]^2\times\mathbb{S}^{d-2}}$ corresponds to a unique point of ${\Omega}$. Thus, there is a bijective mapping as ${\mathscr{F}_s:\Omega\rightarrow{[0,1]^2\times\mathbb{S}^{d-2}}}$ that maps the original non-convex space ${\Omega}$ onto a Euclidean convex transformed space which is a unit tube (because $[0,1]^2\times\mathbb{S}^{d-2}$ can be seen as $[0,1]\times\mathbb{B}^{d-1}$). \Cref{fig:IntrinsicMean} (left) depicts a distribution of points inside a 2D generalized tube (not a trumpet). The non-intrinsic Euclidean mean is invalid as it lies outside the region. \Cref{fig:IntrinsicMean} (middle) illustrates the mapped distribution based on the skeletal coordinate system of the region to the unit tube. \Cref{fig:IntrinsicMean} (right) shows the intrinsic mean as ${\mathscr{F}_s^{-1}(\mu)}$, where ${\mu}$ is the Euclidean mean of the mapped distribution.\par
For a space, such as a trumpet, that can be seen as a generalized tube with an infinitely long spine, we assume ${\lambda\in\mathbb{R}^+}$. Hence, the mapping defines the coordinate system corresponding to the infinite tube as ${\mathscr{F}_s:\Omega\rightarrow{[0,1]\times\mathbb{S}^{d-2}\times\mathbb{R}^+}}$. By taking each trumpet as $\Omega$, the mapping $\mathscr{F}_s$ maps trumpets into an infinite tube, as depicted in \Cref{fig:hypertrumpet} (left).\par
In other words, since the spine of a trumpet is a straight line, the Frenet frame can be considered the identity frame. Thus, by representing $(\bm{v},x,r)$ as ${(\|\bm{v}\|,\bm{u},x,r)}$, where ${\bm{u}=\frac{\bm{v}}{\|\bm{v}\|}}$ is the direction of $\bm{v}$, the mapping ${\mathscr{F}_s:\mathcal{A}_c\to \arc{\mathcal{A}}_c}$ defines a localized mapping that operates on the first component of ${(\|\bm{v}\|,\bm{u},x,r)}$ (i.e., the size of $\bm{v}$, without altering the other components) as ${\mathscr{F}_s((\|\bm{v}\|,\bm{u},x,r)) = (\varsigma,\bm{u}, x, r)}$, where ${\varsigma = \frac{\|\bm{v}\|}{\min\{1, \frac{x}{r}\}}}$, and ${\arc{\mathcal{A}}_c=[0,1]\times\mathbb{S}^{d-2}\times(\mathbb{R}^+)^2}$. Therefore, the mapping $\mathscr{F}_s$ ensures that each element of $\arc{\mathcal{A}}_c$ corresponds to an element of $\mathcal{A}$ representing a valid circular cross-section. Similarly, applying the localized mapping on elements of $\mathcal{A}$, we have ${\mathscr{F}_s((\|\bm{v}\|,\bm{u}, \theta, x, a, b)) = (\varsigma,\bm{u}, \theta, x, a, b)}$. Thus, $\mathscr{F}_s:\mathcal{A}\to \arc{\mathcal{A}}$ defines a bijective mapping between the non-convex region $\mathcal{A}$ and the Euclidean convex region in $\mathbb{R}^6$ (i.e., ${\arc{\mathcal{A}}=\mathbb{B}^2\times[-\pi,\pi]\times(\mathbb{R}^+)^3}$), ensuring that each element of $\arc{\mathcal{A}}$ corresponds to an element of $\mathcal{A}$ representing a valid elliptical cross-section.\par
Intuitively speaking, the hypertrumpet is a symmetric space with a skeletal structure formed by radial vectors with tails at ${(0,\bm{u},\theta,x,a,b)}$ and tip at ${(\min\{1,\frac{x}{r}\},\bm{u},\theta,x,a,b)}$. In this sense, $\mathscr{F}_s$ provides an intrinsic coordinate system for the hypertrumpet. We extend this intrinsic coordinate system to the \textit{poly-hypertrumpets} (i.e., a Cartesian product of multiple hypertrumpets) ${\mathcal{A}^{n+1}}$ based on the coordinate system of the hypertrumpet ${\mathcal{A}}$, while the mapped space ${(\mathscr{F}_s(\mathcal{A}))^{n+1}}$ represents the Euclidean convex transformed space ${\arc{\mathcal{A}}^{n+1}}$. Also, ${\mathcal{A}^{n+1}}$ serves as a space for ETReps representing 2D e-tubes (where $\forall i$: ${b_i\approx 0}$ and ${\psi_i = 0}$).\par
In this work, as implemented in the ETRep R package \citep{ETRep2024}, we prefer to represent cross-sections using the roll angle format $(\bm{v}, \psi, x, a, b)$ rather than the twisting angle format $(\bm{v}, \theta, x, a, b)$. This preference arises because when $\|\bm{v}\| \to 0$ (i.e., $\bm{t}_{i} \to \bm{t}_{i-1}$), oscillations in the normal vector are introduced \citep{farouki2016rational}. These oscillations require frequent adjustments to the resulting frame, which can bias the statistical analysis and lead to unnecessary computational overhead during implementation.\par
Based on the defined ETRep space, we are in a position to define the path, distance, and mean. Assume ETReps ${s_1=(\bm{\omega}_{1i})_{i=0}^n}$ and ${s_2=(\bm{\omega}_{2i})_{i=0}^n}$ in ${\mathcal{A}^{n+1}}$. We define the intrinsic path between ${s_1}$ and ${s_2}$ as ${(\mathscr{F}_s^{-1}(\zeta(\gamma,\mathscr{F}_s(\bm{\omega}_{1i}),\mathscr{F}_s(\bm{\omega}_{2i}))))_{i=0}^n}$, as depicted in \Cref{fig:transformation3D} (bottom row). Let ${s_1,...,s_m}$ be a set of ETReps. By assuming the intrinsic distance as the Euclidean distance in ${\arc{\mathcal{A}}^{n+1}}$, the intrinsic mean (as illustrated in \Cref{fig:selfIntersectionHippo} (c)) is the inverse map of the Euclidean mean of the mapped ETReps as

\begin{equation}\label{equ:meanIntrinsic}
    \bar{s}=\mathscr{F}_s^{-1}\left(\frac{1}{m}\sum_{j=1}^m\mathscr{F}_s(s_j)\right).
\end{equation}
\subsection{Shape analysis}\label{sec:shape_analysis}
In the previous section, we defined the space for unscaled ETReps as the size-and-shape space. In this section, we eliminate, in addition, scaling from the ETReps to define the shape space. Similar to \Cref{sec:Size-and-shape_analysis}, we explore both non-intrinsic and intrinsic approaches.
\subsubsection{Non-intrinsic approach}\label{sec:shapeAnalysisNonIntrinsic}
In \Cref{sec:Size_and_shape_nonIntrinsic} we represented an ETRep as  ${s=((\bm{f}^*_i,x_i,a_i,b_i))_{i=0}^n}$ in the size-and-shape space ${\mathcal{S}^{n+1}}$. Alternatively, the ETRep can be represented as the tuple ${{s=(\bm{f}^*_{i=0,\ldots,n},x_{i=0,\ldots,n},a_{i=0,\ldots,n},b_{i=0,\ldots,n})=(\bm{f}^*_i,x_i,a_i,b_i)_{i=0}^n}}\in{(\mathbb{S}^3)^{n+1}\times(\mathbb{R}^+)^{3(n+1)}}$. Since the product of sets is commutative up to isomorphism (i.e., the order of the spaces in the product does not affect the underlying structure of the product space),  ${(\mathbb{S}^3)^{n+1}\times(\mathbb{R}^+)^{3(n+1)}}$ is the same as ${\mathcal{S}^{n+1}=(\mathbb{S}^3\times(\mathbb{R}^+)^3)^{n+1}}$ (as defined in \Cref{sec:Size_and_shape_nonIntrinsic}). Thus, we can consider ${\mathcal{S}^{n+1}=(\mathbb{S}^3)^{n+1}\times(\mathbb{R}^+)^{3(n+1)}}$.\par
Obviously, the act of uniform scaling does not influence the orientation of the frames. We define the normalized ${s}$ as ${\tilde{s}=(\bm{f}^*_{i},\tilde{x_i},\tilde{a_i},\tilde{b_i})_{i=0}^n}$, where ${\tilde{x_i}=\frac{x_i}{\ell}}$, ${\tilde{a_i}=\frac{a_i}{\ell}}$, ${\tilde{b_i}=\frac{b_i}{\ell}}$ and ${\ell=\norm{(x_i,a_i,b_i)_{i=0}^n}_1}$ is the size of ${s}$, defined by the ${\ell_1}$-norm ${\norm{.}_1}$ (i.e., sum of absolute values).\par
\begin{remark}\label{label:remark1}
The size of a normalized ETRep is 1 \textup{(see the proof in \hyperlink{link_Supplementary}{Supp.Mat})}.
\end{remark}
Therefore, ${\tilde{s}}$ belongs to the shape space ${{\tilde{\mathcal{S}}^{n+1}=(\mathbb{S}^3)^{n+1}\times(\mathbb{H}^+)^{3(n+1)}}\subset\mathcal{S}^{n+1}}$, where ${(\mathbb{H}^+)^{3(n+1)}}$ is the hyperplane restricted to the positive orthant of the Euclidean space, that is, ${{(\mathbb{H}^+)^{3(n+1)}=\{\bm{x}\in(\mathbb{R}^+)^{3(n+1)} \mid \bm{1}\cdot\bm{x}=1\}}}$. Note that although normalizing using the ${\ell_2}$-norm is common, we opt for the ${\ell_1}$-norm. Normalization with the ${\ell_2}$-norm projects the data onto the surface of a curved Riemannian manifold as a hypersphere (analogous to Kendall’s pre-shape space), which introduces additional complexity to the analysis. For example, it complicates controlling conditions like ${\forall i: \bar{a}_i>\bar{b}_i}$ when calculating the mean. We defer the exploration of this more complex scenario to future research.\par
Assume ${m}$ normalized ETReps ${\{\tilde{s}\}_{j=1}^m=\{(\bm{f}^*_{ji},\tilde{x}_{ji},\tilde{a}_{ji},\tilde{b}_{ji})_{i=0}^n\}_{j=1}^m}$. Analogous to \Cref{sec:Size_and_shape_nonIntrinsic}, the transformation path between any two normalized ETReps such as ${\tilde{s}_1}$ and ${\tilde{s}_2}$ can be defined element-wise as ${(\zeta_g(\gamma;\bm{f}^*_{i1},\bm{f}^*_{i2}),\zeta(\gamma;\tilde{x}_{i1},\tilde{x}_{i2}),\zeta(\gamma;\tilde{a}_{i1},\tilde{a}_{i2}),\zeta(\gamma;\tilde{b}_{i1},\tilde{b}_{i2}))_{i=0}^n}$ with \Cref{equ:non-intrinsic_distance} defining the distance. Therefore, based on \Cref{equ:non-intrinsic_mean}, the non-intrinsic mean shape is ${{\bar{\tilde{s}}_{\text{non.in}}=(\bar{\bm{f}}^*_{i},\bar{\tilde{x}}_i,\bar{\tilde{a}}_i,\bar{\tilde{b}}_i)_{i=0}^n}}$, where ${\forall i}$: ${\bar{\bm{f}}^*_i}$ is the mean frame of ${\{\bm{f}^*_{ji}\}_{j=1}^m}$, and ${\bar{\tilde{x}}_i}$, ${\bar{\tilde{a}}_i}$, and ${\bar{\tilde{b}}_i}$ are the arithmetic means of ${\{\tilde{x}_{ji}\}_{j=1}^m}$, ${\{\tilde{a}_{ji}\}_{j=1}^m}$, and ${\{\tilde{b}_{ji}\}_{j=1}^m}$, respectively. The size of ${\bar{\tilde{s}}_{\text{non.in}}}$ is clearly 1.0 as it lies within ${\tilde{\mathcal{S}}^{n+1}}$.
\subsubsection{Intrinsic approach}\label{sec:shape_analysis_Intrinsic} 
Following \Cref{sec:shapeAnalysisNonIntrinsic}, we can represent ETRep ${s=((\bm{v}_{i},\psi_i,x_i,a_i,b_i))_{i=0}^n}$ as a vector ${\bm{s}=(\bm{v}_{i},\psi_i,x_i,a_i,b_i)_{i=0}^n}$ in the actual space ${\mathcal{A}^{n+1}\subset(\mathbb{B}^2)^{n+1}\times[-\pi,\pi]^{n+1}\times(\mathbb{R}^+)^{3(n+1)}}$. Again, the act of scaling does not affect ${\bm{v}_i}$ and ${\psi_i}$ defining the frames' orientations. Thus, we consider the normalized ${\bm{s}}$ as ${\tilde{\bm{s}}=(\bm{v}_{i},\psi_i,\frac{x_i}{\ell},\frac{a_i}{\ell},\frac{b_i}{\ell})_{i=0}^n}$, where ${\ell}$ is the size of ${s}$.\par
\begin{remark}\label{label:remark2}
Uniform scaling does not affect the RCC \textup{(see the proof in \hyperlink{link_Supplementary}{Supp.Mat})}.
\end{remark}
As outlined in \Cref{sec:size_and_shape_analysis_Intrinsic}, the mapping ${\mathscr{F}_s}$ preserves the value of ${x_i}$, ${a_i}$, and ${b_i}$. Thus, ${\mathscr{F}_s(\tilde{\bm{s}})}$ can be seen as a vector as ${(\varsigma_i\bm{u}_{i},\psi_i,\frac{x_i}{\ell},\frac{a_i}{\ell},\frac{b_i}{\ell})_{i=0}^n}$ in the Euclidean convex space ${(\mathbb{B}^2)^{n+1}\times[-\pi,\pi]^{n+1}\times(\mathbb{H}^+)^{3(n+1)}\subset\arc{\mathcal{A}}^{n+1}}$. Consequently, the path, transformation, and distance can be defined in the Euclidean sense within the space ${(\mathbb{B}^2)^{n+1}\times[-\pi,\pi]^{n+1}\times(\mathbb{H}^+)^{3(n+1)}}$. \hyperlink{link_Supplementary}{Supp.Mat.} provides an example illustrating the intrinsic and non-intrinsic transformations between e-tubes, both with and without scaling. In the example, the non-intrinsic method results in invalid outcomes due to self-intersections.\par
\begin{figure}[ht]
\centering
\boxed{\includegraphics[width=0.70\textwidth]{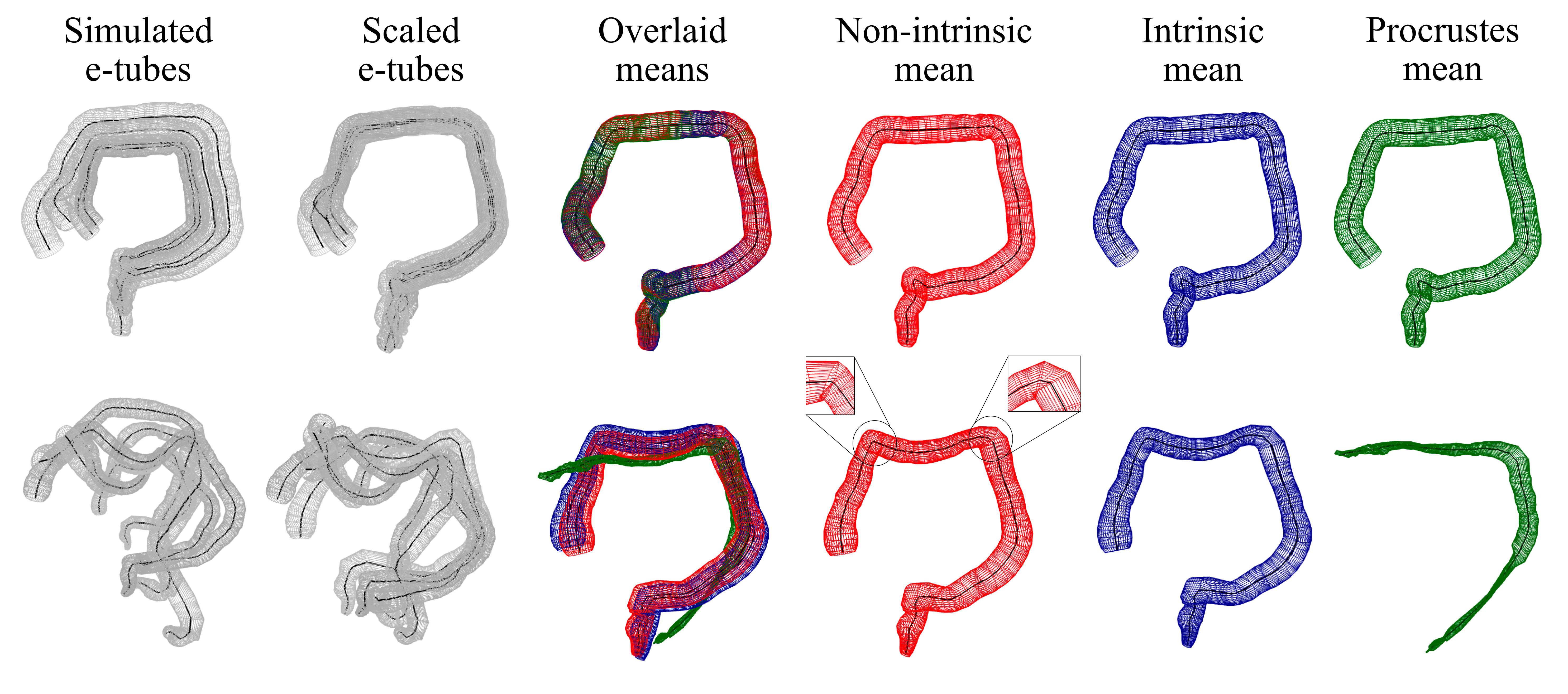}}
\caption{Comparison of the means for two examples (displayed in two rows) from a set of e-tubes simulated based on a colon-shaped e-tube. The objects in the two left columns are overlaid using the GPA. First row: All mean shapes are valid. Second row: The non-intrinsic mean is invalid due to violations of the RCC at two locations, highlighted by circles. The intrinsic mean is valid. The Procrustes mean is invalid due to collapsing.}
\label{fig:MeanShapes}
\end{figure}
Consequently, given a set of normalized ETReps ${\tilde{\bm{s}}_1,...,\tilde{\bm{s}}_m}$ in ${\mathcal{A}^{n+1}}$, based on \Cref{equ:meanIntrinsic}, the mean is ${\bar{\tilde{\bm{s}}}=\mathscr{F}_s^{-1}(\frac{1}{m}\sum_{j=1}^m\mathscr{F}_s(\tilde{\bm{s}}_j))}$. Clearly, the size of $\bar{\tilde{\bm{s}}}$ is 1, as it resides in ${(\mathbb{B}^2)^{n+1}\times[-\pi,\pi]^{n+1}\times(\mathbb{H}^+)^{3(n+1)}}$. \Cref{fig:MeanShapes} compares the intrinsic mean, extrinsic mean, and Procrustes mean of a set of e-tubes simulated based on an e-tube model of a colon (from \Cref{fig:hippoMandibleColon}). The simulation introduces variations in ${\bm{v}_i}$, ${\psi_i}$, and the size of the reference model, as detailed in \hyperlink{link_Supplementary}{Supp.Mat}. The superimposition of objects and the Procrustes mean are computed using \textit{Generalized Procrustes Analysis} (GPA) \citep{dryden2016statistical}. In the second column (from the left), the objects are scaled according to the defined e-tube size. In the first row, the means are valid and nearly identical, reflecting the small variation between the simulated samples. As the variation increases, as shown in the second row, the differences between the means become more pronounced. The intrinsic mean remains valid, whereas the non-intrinsic mean becomes invalid due to violations of the RCC at two locations. Additionally, the Procrustes mean is invalid due to collapsing. 
\section{Application}\label{sec:Application}
A key application of sample mean shape analysis is in investigating shape differences through hypothesis testing. This section explores ETRep hypothesis testing using the intrinsic shape analysis outlined in \Cref{sec:size_and_shape_analysis_Intrinsic,sec:shape_analysis_Intrinsic}. Furthermore, in \Cref{sec:Real_data_analysis}, we apply the proposed hypothesis testing methods to study a real hippocampal dataset from the ParkWest study \citep{alves2009incidence} provided by Stavanger University Hospital.\par
As discussed in \Cref{sec:ETRep_Analysis}, an ETRep (or a scaled ETRep) can be seen as a ${6(n+1)}$D vector in the Euclidean convex transformed space ${\arc{\mathcal{A}}^{n+1}\subset\mathbb{R}^{6(n+1)}}$ as ${(\varsigma_i\bm{u}_{i},\psi_i,x_i,a_i,b_i)_{i=0}^n}$. Let ${A=\{\tilde{s}_j^{A}\}_{j=1}^{m_1}}$ and ${B=\{\tilde{s}_j^{B}\}_{j=1}^{m_2}}$ be two groups of ETReps. Thus, we have two distinct ${6(n+1)}$D multivariate distributions in $\arc{\mathcal{A}}^{n+1}$. The global test can be considered as ${{H_{0}:\mu_{A}=\mu_{B}}}$ versus ${{H_{1}:\mu_{A}\neq\mu_{B}}}$ where ${\mu_{A}}$ and ${\mu_{B}}$ are the observed Euclidean sample means in the feature space $\arc{\mathcal{A}}^{n+1}$. For the hypothesis testing, we consider the permutation test with minimal assumptions. Given the pooled group of two data sets (here ${A\cup B}$), the permutation method randomly partitions the pooled group into two groups of sizes ${m_1}$ and ${m_2}$ without replacement many times and measures the test statistics between the paired groups. The empirical ${p}$-value is ${({1+\sum_{h=1}^{\text{N}}\mathds{1}(|t_{h}|\geq{|t_{o}|})})/(\text{N}+1)}$, where ${t_{o}}$ is the observed test statistic (e.g., Euclidean distance), ${t_{h}}$ is the ${h}$th permutation test statistic, ${\text{N}}$ is the number of permutations (usually greater than ${10^4}$), and $\mathds{1}$ is the indicator function (i.e., ${\mathds{1}(X)=1}$ if ${X}$ is true, otherwise ${\mathds{1}(X)=0}$). Since the feature space is high-dimensional, we chose the \textit{direction projection permutation} (DiProPerm) approach \citep{wei2016direction}.\par
DiProPerm evaluates the separation between two groups by projecting high-dimensional data onto a vector that maximizes their distinction and then testing this separation's significance through random permutations of group labels. Each data point $\bm{x}_i$ is projected onto the mean difference vector $\bm{d}= \bar{\bm{x}} - \bar{\bm{y}}$, yielding scalar projections $z_i = \bm{x}_i^\top \bm{d}$ for each group. The group means, $\bar{z}_1$ and $\bar{z}_2$, are compared to quantify the difference.\par
To detect local dissimilarities, we perform element-wise comparisons of ETReps using partial permutation tests for each feature, testing ${H_{0k}:\mu_{A}(k)=\mu_{B}(k)}$ against ${H_{1k}:\mu_{A}(k)\neq \mu_{B}(k)}$, where $\mu_{A}(k)$ and $\mu_{B}(k)$ represent the observed sample means for the $k$-th feature and ${k = 1,...,6(n+1)}$. The test statistics is the t-statistic ${t=\frac{\bar{x} -\bar{y}}{\hat{\sigma}_p\sqrt{\frac{1}{m_1}+\frac{1}{m_2}}}}$, where $\hat{\sigma}_p$ is the pooled standard deviation. Further, to control false positives arising from multiple comparisons, we adjust the $p$-values using the approach of \citet{benjamini1995controlling} (BH). The BH method ranks $p$-values and adjusts them upward to account for multiple tests, reducing the likelihood of false positives following $p_{\text{adjusted}}(k) =\frac{p(k)\times K}{k}$, where $p(k)$ is the $k$-th $p$-value, $K$ is the total number of tests, and $k$ is the rank.\par
\subsection{Real data analysis}\label{sec:Real_data_analysis}
In this section, we compare the ETRep of the left hippocampus in patients with Parkinson's disease (PD) to that of a healthy control group (CG). The PD comprises 117 samples, while the CG includes 67 samples. Each hippocampus is represented by an ETRep consisting of 53 cross-sections approximately equally spaced across their spines (based on curve-length registration). The model fitting procedure and further analysis, including SPHARM-PDM and radial distance analysis of \citet{thompson2004mapping}, are detailed in the \hyperlink{link_Supplementary}{Supp.Mat}.\par
\begin{figure}[ht]
\centering
\boxed{\includegraphics[width=0.98\textwidth]{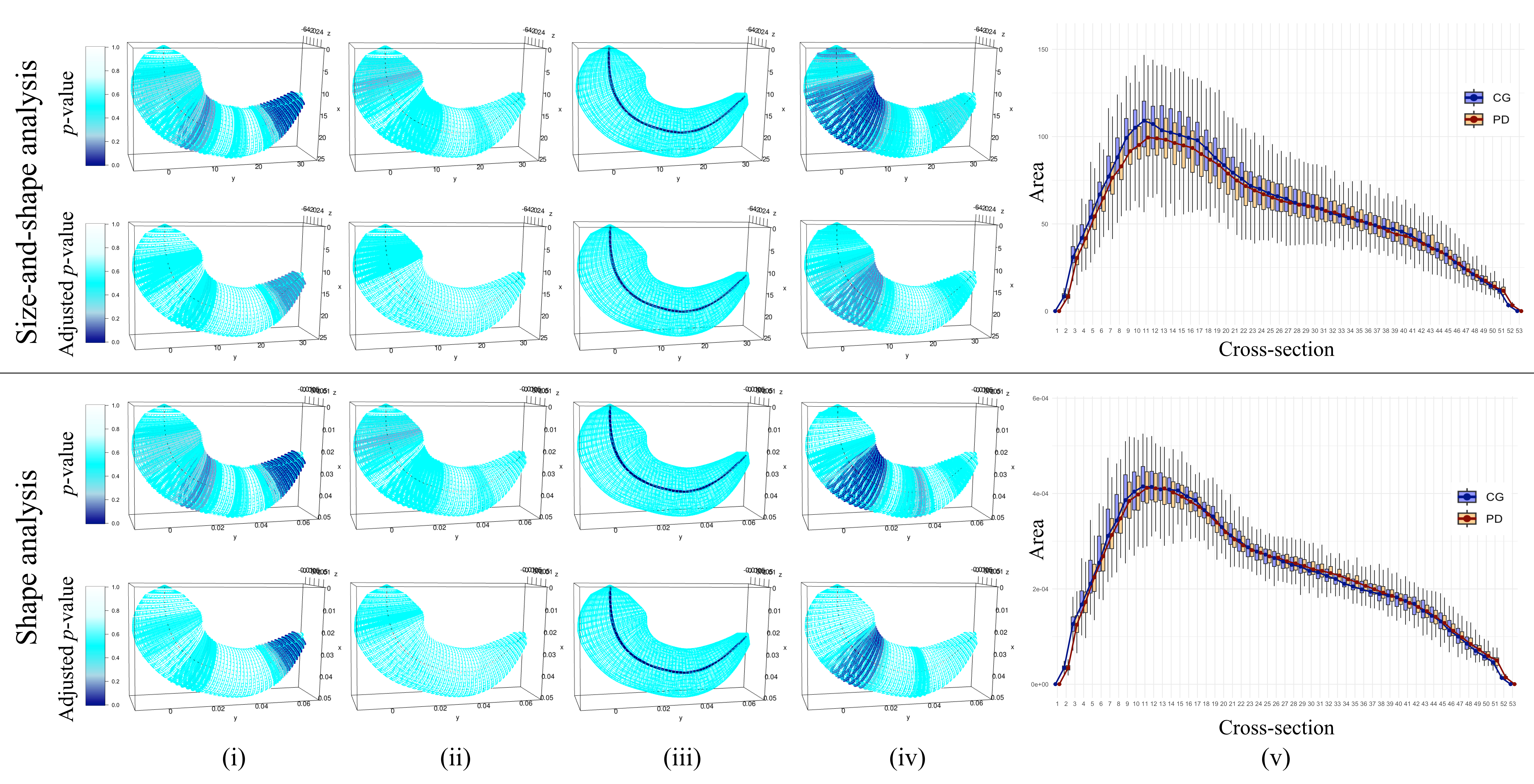}}
\caption{Hypothesis testing on ETReps of CG versus PD based respectively on raw ${p}$-values and adjusted ${p}$-values with and without scaling. Significant features associated with ${\bm{v}}$ vectors, roll angles ${\psi}$, spinal connection lengths ${x}$, and principal radii ${a, b}$ are highlighted in bold across columns (i-iv), respectively. Column (v) compares the area of the corresponding cross-sections.}
\label{fig:hypothesisTest}
\end{figure}
At a significance level of 0.1, the global DiProPerm test indicated a significant difference between the CG and PD groups, with p-values of 0.086 for the scaled data and 0.044 for the original data. Additionally, \Cref{fig:hypothesisTest} highlights significant features from the partial tests, capturing cross-sectional rotation, degree of twist, spinal elongation, and cross-sectional size without $ p $-value adjustment. These features correspond to the vectors $ \mathbf{v} $, roll angles $\psi$, spinal connection lengths $ x $, and principal radii $ a $ and $ b $, as shown in columns (i)-(iv). Column (v) compares the area (i.e., $ \pi a_i b_i $) of the corresponding cross-sections. Evident atrophy in PD is observed in the head of the hippocampi (i.e., the figures' anterior parts associated with the front section of the medial temporal lobe). Other significant differences are associated with the curvature at the tail of the hippocampi (i.e., the figures' posterior parts that extend toward the back of the brain) and the spinal length.
\section{Conclusion}\label{sec:Conclusion}
In the existing literature, many prevalent shape spaces overlook the inherent geometric structure of their underlying shape representations, such as PDMs, leading to the inclusion of invalid shapes in the statistical analysis. In this work, we proposed a novel framework for statistically analyzing swept regions, specifically addressing the challenge of avoidance of self-intersection. We focused on a particular class of swept regions known as elliptical tubes (e-tubes). We introduced the e-tube representation (ETRep) as a robust method for representing e-tubes. The ETRep was designed to be invariant to rigid transformations and could be described as a sequence of elliptical cross-sections along the object's central curve. To tackle the issue of self-intersection in ETRep analysis, we discussed the relative curvature condition (RCC). By incorporating the RCC, we defined ETRep size-and-shape space and shape space equipped with an intrinsic skeletal coordinate system. We described the ETRep intrinsic distance and mean shape within the defined spaces. We proposed an intrinsic distance measure for the underlying spaces to prevent self-intersection and ensure valid statistical analysis. In \hyperlink{link_Supplementary}{Supp.Mat.}, we have also briefly addressed the problem of non-local intersections and suggested a solution for a specific class of ETReps that are simply straightenable. Further development of this solution could serve as a promising direction for future research. Finally, we demonstrated the application of ETRep analysis by comparing the hippocampi of patients with Parkinson’s disease to those of a healthy control group. The ETRep framework provided detailed insights, effectively identifying the types and precise structural differences between the groups. To complement this study, the ``ETRep'' R package \citep{ETRep2024} is provided to facilitate the application of the discussed methodologies and support the methods and analysis presented.

\section*{Acknowledgments}
This work is funded by the Department of Mathematics and Physics of the University of Stavanger. We extend our gratitude to Prof. J. S. Marron for his invaluable guidance and to the late Prof. James Damon for the inspiration for this work. We also thank Prof. Guido Alves and Prof. Ole-Bjørn Tysnes for generously providing the ParkWest data.
\begin{center}
{\large\bf Supplementary Materials}
\end{center}
\textbf{Supplementary:} \hypertarget{link_Supplementary} Supplementary Materials referenced in this work are available as a pdf.
\textbf{R-code:} In \texttt{Supplementary\_Materials.zip}, R codes and files are placed. (zip)

% \end{itemize}
% \bibliographystyle{unsrtnat}
% \bibliographystyle{unsrt}
\bibliographystyle{spbasic}
\bibliography{myBibFile}

\end{document}